\documentclass[sigconf]{acmart}

\copyrightyear{2019}
\acmYear{2019}
\setcopyright{iw3c2w3}
\acmConference[WWW '19]{Proceedings of the 2019 World Wide Web Conference}{May 13--17, 2019}{San Francisco, CA, USA}
\acmBooktitle{Proceedings of the 2019 World Wide Web Conference (WWW '19), May 13--17, 2019, San Francisco, CA, USA}
\acmPrice{}
\acmDOI{}
\acmISBN{}

\usepackage{booktabs}
\usepackage{url}
\usepackage{xcolor}
\usepackage{graphicx}
\usepackage{amssymb,amsmath}
\usepackage{caption}
\usepackage{subcaption}
\usepackage{float}
\usepackage{balance}

\newcommand{\hashtag}{h}
\newcommand{\temporal}{\mathcal{T}}
\newcommand{\Hashtag}{\mathcal{H}}
\newcommand{\vecmat}{M}
\newcommand{\paramat}{X}
\DeclareMathOperator*{\argmin}{arg\,min}
\DeclareMathOperator*{\argmax}{arg\,max}
\newcommand{\user}{u}
\newcommand{\User}{\mathcal{U}}

\usepackage{balance}

\begin{document}

\title{Language in Our Time: An Empirical Analysis of Hashtags}

\author{Yang Zhang}
\affiliation{%
  \institution{CISPA Helmholtz Center for Information Security\\
  Saarland Informatics Campus}
}
\email{yang.zhang@cispa.saarland}

\renewcommand{\shortauthors}{Yang Zhang}

\begin{abstract}
Hashtags in online social networks have gained tremendous popularity during the past five years. The resulting large quantity of data has provided a new lens into modern society. Previously, researchers mainly rely on data collected from Twitter to study either a certain type of hashtags or a certain property of hashtags. In this paper, we perform the first large-scale empirical analysis of hashtags shared on Instagram, the major platform for hashtag-sharing. We study hashtags from three different dimensions including the temporal-spatial dimension, the semantic dimension, and the social dimension. Extensive experiments performed on three large-scale datasets with more than 7 million hashtags in total provide a series of interesting observations. First, we show that the temporal patterns of hashtags can be categorized into four different clusters, and people tend to share fewer hashtags at certain places and more hashtags at others. Second, we observe that a non-negligible proportion of hashtags exhibit large semantic displacement. We demonstrate hashtags that are more uniformly shared among users, as quantified by the proposed hashtag entropy, are less prone to semantic displacement. In the end, we propose a bipartite graph embedding model to summarize users' hashtag profiles, and rely on these profiles to perform friendship prediction. Evaluation results show that our approach achieves an effective prediction with AUC (area under the ROC curve) above 0.8 which demonstrates the strong social signals possessed in hashtags.
\end{abstract}

%
%

\ccsdesc[500]{Information systems~Data mining}
\ccsdesc[500]{Human-centered computing~Social networks}
\ccsdesc[500]{Human-centered computing~Social tagging}
\ccsdesc[500]{Human-centered computing~Social media}
\ccsdesc[500]{Human-centered computing~Empirical studies in collaborative and social computing}

\keywords{Hashtag, online social networks, data analysis}

\maketitle

\section{Introduction}
\label{sec:intro}

\begin{figure}[!t]
\centering
\includegraphics[width=\columnwidth]{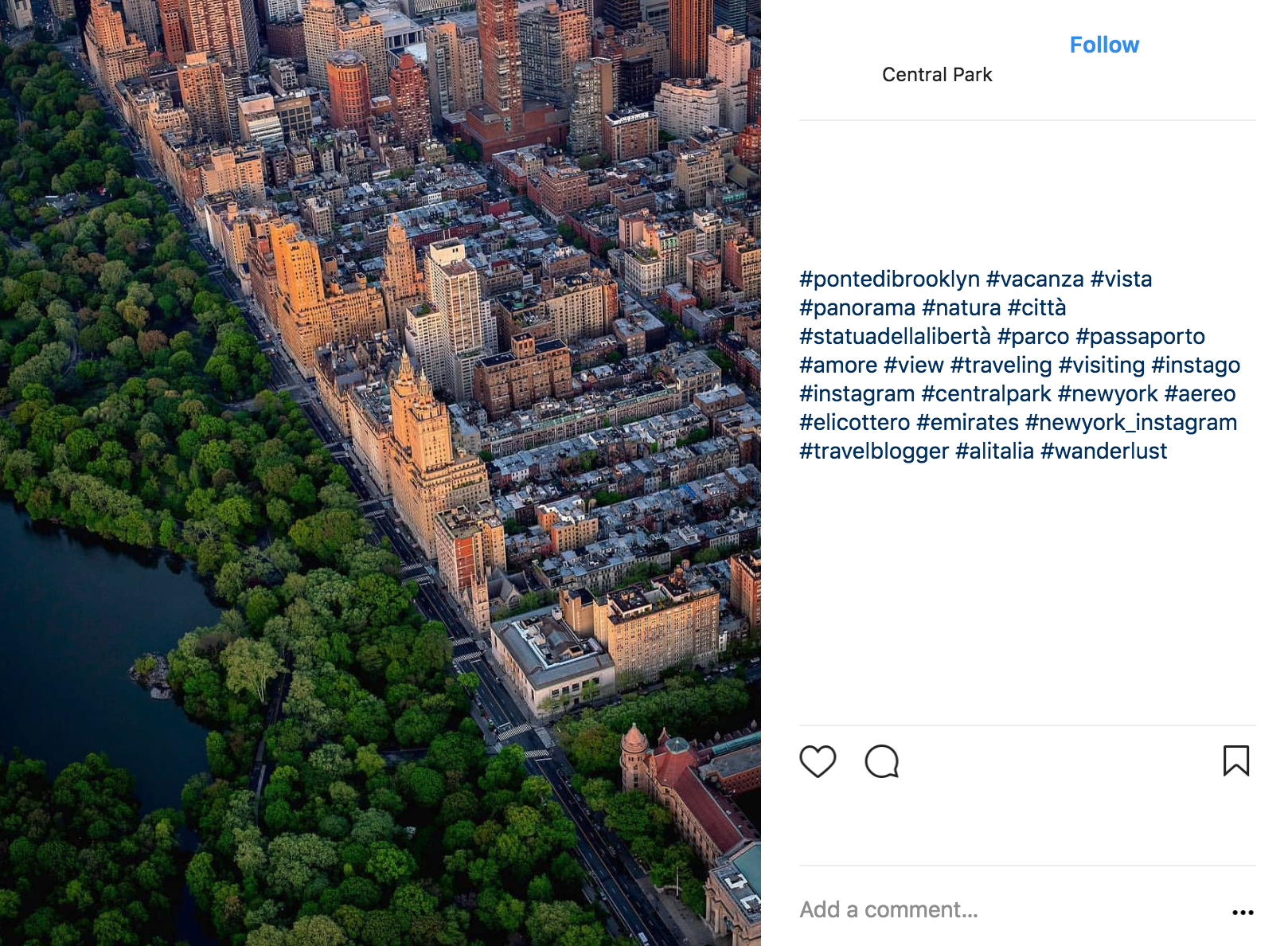}
\caption{An Instagram post with multiple hashtags.}
\label{fig:example-instagram}
\end{figure} 

The last decade has witnessed 
the explosive development of online social networks (OSNs).
Leading players in the business, 
such as Facebook,\footnote{\url{https://www.facebook.com/}} Twitter,\footnote{\url{https://twitter.com/}} 
and Instagram,\footnote{\url{https://www.instagram.com/}}
have become the major platform for people to share life moments,
communicate with each other, and maintain social relations.
Moreover, OSNs have introduced many new notions into our society,
such as ``like'', ``share'', and ``check-in''.
One particular interesting notion of this kind is \emph{hashtag}.

Created back in 2007,
hashtags are designed to help users efficiently retrieve information on Twitter.
With development,
Instagram has become the major platform for hashtag-sharing.
Nowadays, it is very common to see an Instagram post 
associated with multiple hashtags (see Figure~\ref{fig:example-instagram} for an example).
People also start to use hashtags for various purposes.
For instance, many brands use hashtags
to promote their products, such as \#mycalvins and \#shareacoke.
Also, hashtags play a major role
in various political movements, e.g., \#blacklivesmatter and \#notmypresident.
More interestingly, hashtags have evolved themselves into a new-era language:
People have created many hashtags the meanings of 
which do not exist in the natural language.
For instance, 
\#nomakeup attached to a photo
indicates that the person in the photo did not wear any makeup;
a user publishing \#follow4follow means 
she will follow back others who follow her in OSNs.
In another example,
\#tbt (throwback Thursday)
indicates the corresponding photo was taken from old days.

The large quantity of hashtags has provided us with a new way 
to understand the modern society.
Previously, researchers have studied hashtags 
from various angles~\cite{RMK11,TR12,SCFYCQA15,FPS15,
MAH16,AW16,OWG16,OAWHT17,
MBLK18,ZHRLPB18}.
For instance, Souza et al.\ have analyzed \#selfie
to understand the online self-portrait convention~\cite{SCFYCQA15}.
Mejova et al.\ use \#foodporn to study people's dining preference
on a global scale~\cite{MAH16}.
More recently, Zhang et al.\ investigate the (location) privacy risks
stemming from sharing hashtags~\cite{ZHRLPB18}.

Most of these previous works
have studied either a certain type of hashtags~\cite{SCFYCQA15,
MAH16,OWG16,OAWHT17,MBLK18} 
or a certain property of hashtags~\cite{RMK11,AW16,ZHRLPB18}.
Meanwhile, several general analyses
concentrate on hashtags shared on Twitter~\cite{TR12,FPS15}
which is a very different OSN from Instagram 
with respect to user group, popularity, and functionality~\cite{HMK14,MHK14}.

In this paper, we perform the first large-scale empirical study
aiming at understanding hashtags shared on Instagram.
Our analyses are centered around three research questions
summarized from three different dimensions, 
i.e., the temporal-spatial dimension, the semantic dimension,
and the social dimension.

\subsection{Research Questions}

Different hashtags exhibit different temporal patterns.
Holiday-related hashtags, such as \#newyear, 
may have periodic popularity,
while the usage of some other hashtags 
may increase steadily over time.
Besides, the information from the spatial dimension 
may also influence users' hashtag-sharing behavior:
People are more (less) willing to share hashtags
when they are at certain places.
Therefore, we ask our first research question:

\medskip
\emph{\textbf{RQ1.} What are the temporal and spatial patterns of hashtags?}
\medskip

As a new-era language,
the semantics of hashtags can uncover 
many underlying patterns of the modern-style communication.
Moreover, due to their inherent dynamic nature,
some hashtags may change their meanings within a short time.
We therefore ask:

\medskip
\emph{\textbf{RQ2.} Do hashtags exhibit semantic displacement?}
\medskip

Following social homophily theory,
we hypothesize that 
friends exhibit more similar hashtag-sharing behavior than strangers.
In another way, hashtags possess strong signals 
for inferring users' social relations.
To test this hypothesis, we ask:

\medskip
\emph{\textbf{RQ3.} Can hashtags be used to infer social relations?}
\medskip

\subsection{Contribution}

We perform the first large-scale empirical analysis of 
hashtags shared on Instagram.
We first sample in total 51,527 Instagram users 
from three major cities in the English-speaking world,
including New York, Los Angeles, and London.
We then collect all the posts 
that these users have shared from the end of 2010 to the end of 2015,
and build three separate datasets.
In total, our datasets contain more than 41 million Instagram posts
shared together with 7 million hashtags.

To address \textbf{RQ1},
we first perform clustering to summarize hashtags' temporal patterns
which results in four different clusters.
In particular, one cluster of hashtags exhibits strong periodic popularity,
some examples in this cluster are \#snow, \#bbq, and \#superbowl.
For the spatial dimension,
we show that people in all the three datasets 
tend to share fewer hashtags at certain types of places,
such as bars,
while more at other types of places, e.g., parks.

For \textbf{RQ2}, we first adopt
the skip-gram model~\cite{MCCD13,MSCCD13} 
to map hashtags to continuous vectors
and demonstrate that these vectors can very well represent hashtags' semantics.
Relying on the orthogonal Procrustes approach, 
we measure a hashtag's semantic displacement between two consecutive years
as the distance between its two vectors trained at those years.
Evaluation shows that a non-negligible proportion (more than 10\%) of hashtags 
indeed shift their meanings to a large extent.
We further define a notion, namely hashtag entropy,
to quantify the uniformity of hashtags being shared among users,
and observe that hashtags with low entropy
are prone to semantic displacement:
Correlation coefficients between semantic displacement and entropy 
in all the datasets are below -0.6.

To answer \textbf{RQ3}, we perform a friendship prediction task
solely based on hashtags.
We propose a bipartite graph embedding approach 
to learn each user's hashtag profile
and conduct unsupervised friendship prediction 
based on two users' profiles' cosine distance.
Extensive experiments show that our approach achieves an effective prediction
with AUC (area under the ROC curve) above 0.8 in all the three datasets,
and outperforms several baseline models by 20\%.
This indicates that hashtags indeed possess strong signals on social relations.

\medskip
To the best of our knowledge,
no previous works have studied hashtags' spatial patterns, semantic displacement, and social signals.
We are the first to analyze hashtags from these angles.

We believe our analysis can benefit several parties.
The conclusions drawn from answering \textbf{RQ1} and \textbf{RQ2}
can help media campaigns to design more attractive hashtags to engage new customers.
The semantic displacement result (from \textbf{RQ2})
can help researchers gain a deeper understanding of the OSN culture.
The friendship prediction algorithm derived from answering \textbf{RQ3}
shows that hashtags can also be used as a strong signal for friendship recommendation,
which is essential for OSN operators.

\subsection{Organization}

The rest of the paper is organized as the following.
We describe our dataset collection methodology 
with some initial analyses in Section~\ref{sec:dataset}.
In Section~\ref{sec:spatialtemporal},
we investigate the temporal and spatial patterns of hashtags.
Section~\ref{sec:semantic} studies the semantic displacement of hashtags.
In Section~\ref{sec:friends}, 
we concentrate on using hashtags to infer social relations.
Section~\ref{sec:related} discusses the related work in the field.
We conclude the paper in~Section~\ref{sec:conclu}.

\section{Datasets and Initial Analyses}
\label{sec:dataset}

In this section, we first describe our data collection methodology,
then perform some initial analyses on hashtags.

\subsection{Datasets Collection}

We resort to Instagram to collect our datasets for experiments.
Launched in October 2010,
Instagram is a social network service concentrating on photo sharing.
By now, it is the second most popular OSN with more than 1 billion monthly 
active users.\footnote{\url{https://instagram-press.com/our-story/}}
Instagram is the major social network for hashtag-sharing (Figure~\ref{fig:example-instagram}),
many popular hashtags 
are strongly related to Instagram itself,
e.g., \#instagood and \#instamood.

We collect our dataset relying on Instagram's public API on April, 2016.
The first step is finding a sample of users~\cite{MPLC13}.
In the literature, there exist multiple methods for this task.
One is relying on the OSN's streaming API~\cite{PMS18}
which Instagram does not provide.
Another way is generating random integers
and query the Instagram API to see whether these numbers 
are valid user IDs~\cite{SCFYCQA15}.
However, this approach needs us to further investigate 
how Instagram users' IDs are distributed.
In this paper,
we instead sample users based on their locations following previous works~\cite{VMCG09,ZHRLPB18}.

\begin{table}[!t]
\centering
\caption{Summary statistics of all the three datasets.}
\label{table:dataset}
\begin{tabular}{l | c | c | c}
\toprule
& New York & Los Angeles & London\\
\midrule
No.\ posts & 20,673,946 & 11,907,967 & 8,640,637\\
No.\ hashtags & 4,095,575 & 3,071,158 &1,702,675\\
No.\ check-ins & 1,609,062 & 883,862 & 586,420\\
No.\ users & 25,735 & 14,687 & 11,105\\
No.\ social links & 82,964 & 36,434 & 12,900\\
\bottomrule
\end{tabular}
\end{table}

We concentrate on three major English-speaking cities 
including New York, Los Angeles, and London.
In the first step, we query the API of Foursquare,\footnote{\url{https://foursquare.com/}}
a location-based social network,
to find all the Foursquare location IDs 
in the three cities.
Then, we use Instagram's API to map all the Foursquare location IDs
to the corresponding Instagram location IDs.\footnote{Instagram's API was connected 
with Foursquare's API until April 20th, 2016 
(\url{https://www.instagram.com/developer/changelog/}).}
Next, we query all the obtained Instagram location IDs to get all users
who have ever shared posts at those locations.
As in previous works,
we further perform some preprocessing 
to filter out the users matching 
any of the following criteria~\cite{CML11,SNM11,PZ15,PZ17,BHPZ17}.
\begin{itemize}
\item users with less than 20 check-ins in each city
\item users whose numbers of followers are above the 90th percentile (celebrities)
or below the 10th percentile (bots)
\item users not using human images in their profile photos\footnote{This is done with the help of Face++'s API.}
\end{itemize}
In total, we obtain 51,527 Instagram user IDs.
Then, we collect all these users' Instagram posts 
from the creation time of their accounts until December 31st, 2015.
Each post is organized in the following format.
\[
\langle \text{user ID}, \text{time}, \text{hashtags}, \text{location ID}\rangle
\]
Note that a post is not necessarily associated 
with a set of hashtags or a certain location ID.
We further query Instagram's API
to extract users' social relationships.
We consider two users to be friends 
if they follow each other~\cite{CML11,DTWTCRC12,BHPZ17}.

We treat datasets collected from users in the three cities separately
to ensure the robustness of our analyses.
In total, the New York dataset (dataset collected 
from users sampled in New York)
contains more than 20 million Instagram posts,
the Los Angeles dataset contains 11 million posts,
and the London dataset contains 8.6 million posts.
Moreover, the three datasets contain more than 7 million hashtags.
Table~\ref{table:dataset} presents some summary statistics of the datasets.

\begin{figure}
\centering
\includegraphics[width=0.95\columnwidth]{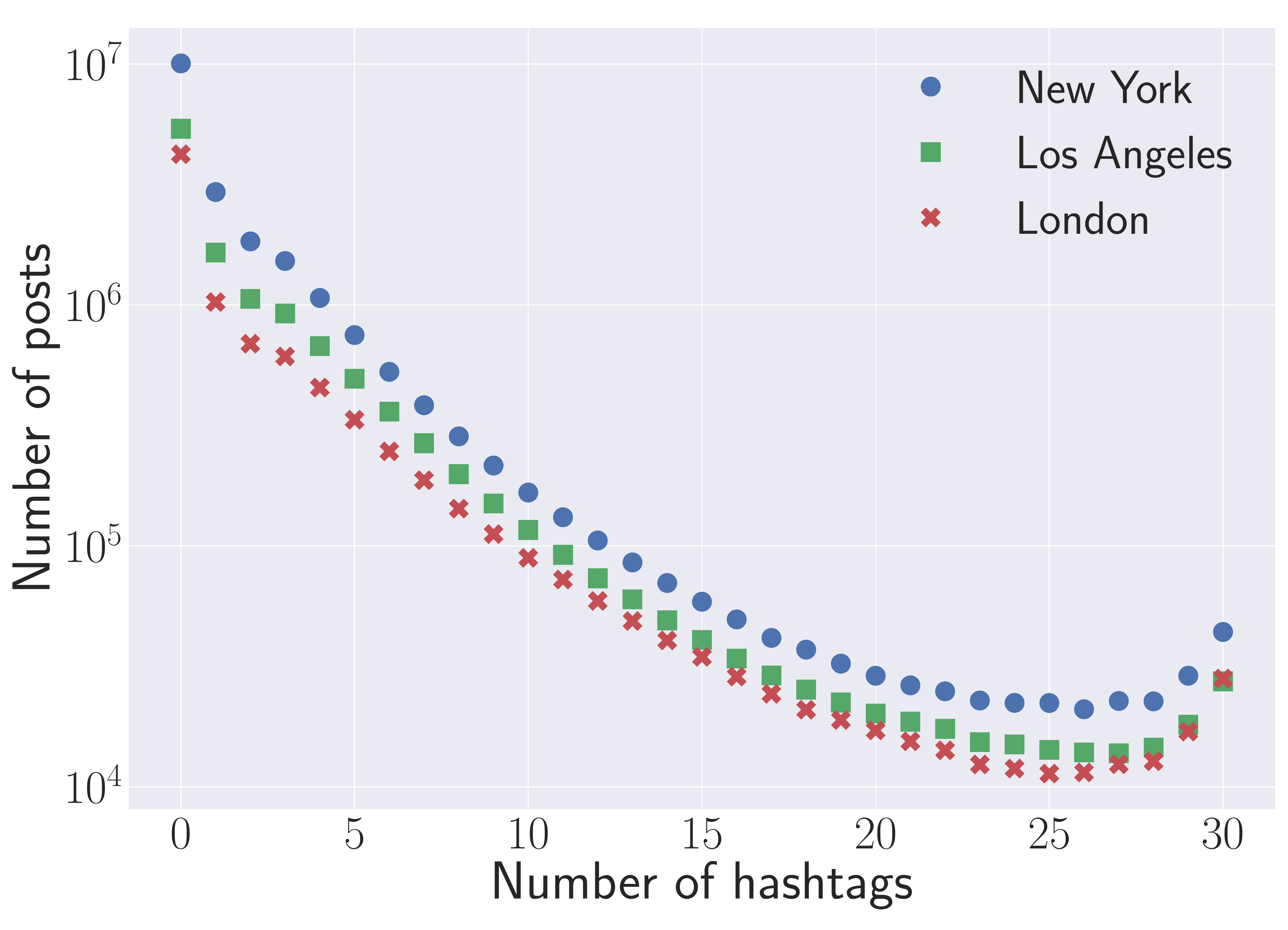}
\caption{Distribution of the number of hashtags in each post in three datasets.
The y-axis is in log scale.
The proportions of posts with no hashtags are 48.69\% in the New York dataset,
45.28\% in the Los Angeles dataset, and 48.92\% in the London dataset.}
\label{fig:media_dist}
\end{figure}

\medskip
\noindent\textbf{Ethical Considerations.}
Our data collection is done through Instagram's public API in 2016.
All the datasets are stored in a central server with restricted access.
We further anonymize the datasets
by removing all users' screen names,
and replacing their Instagram IDs with randomly generated numbers.
Our experiments are conducted on these anonymized datasets.

\begin{figure*}[!t]
\centering
\begin{subfigure}{\columnwidth}
\centering
\includegraphics[width=0.95\columnwidth]{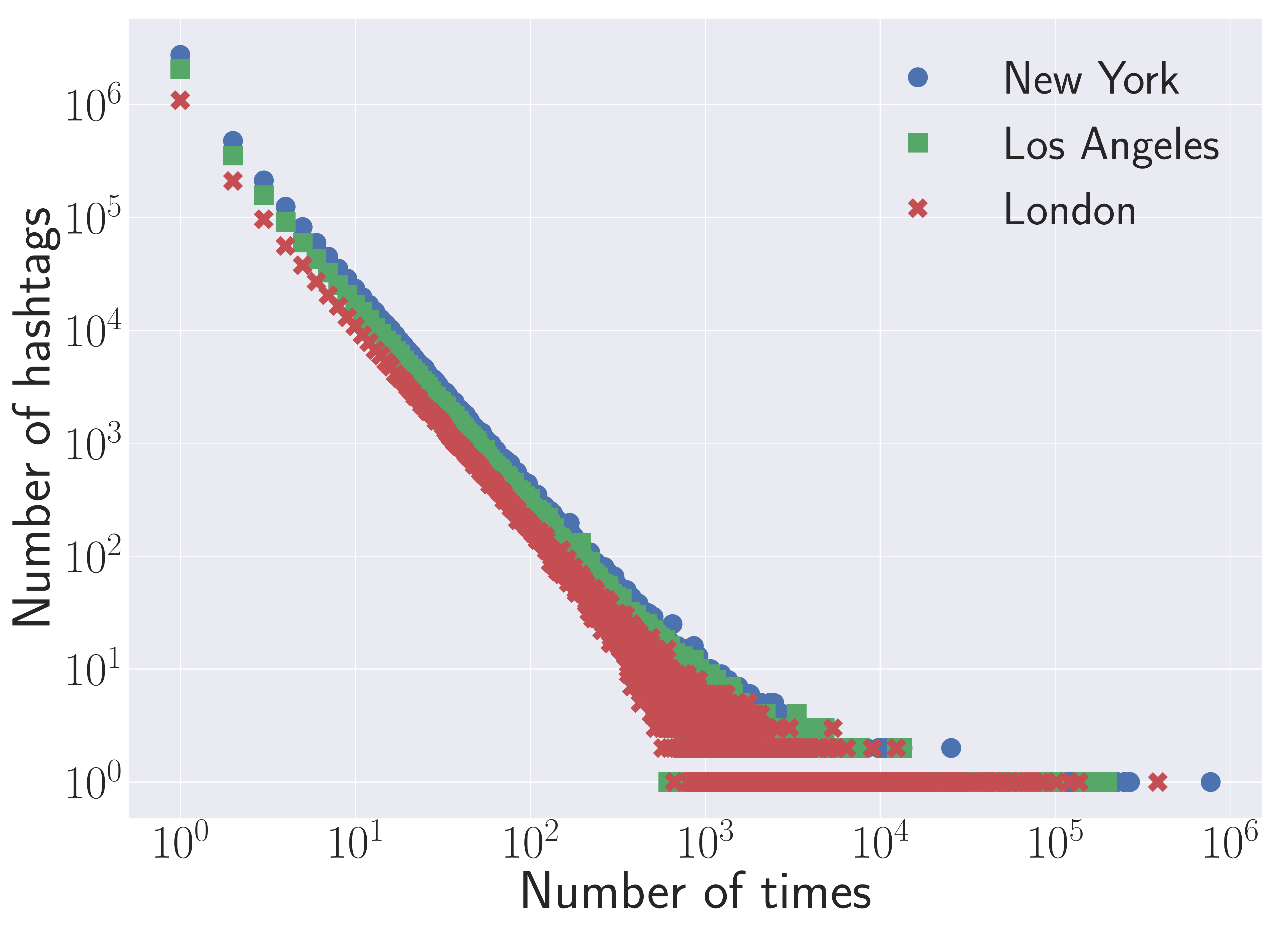}
\caption{}
\label{fig:share_cnt_dist}
\end{subfigure}
\begin{subfigure}{\columnwidth}
\centering
\includegraphics[width=0.95\columnwidth]{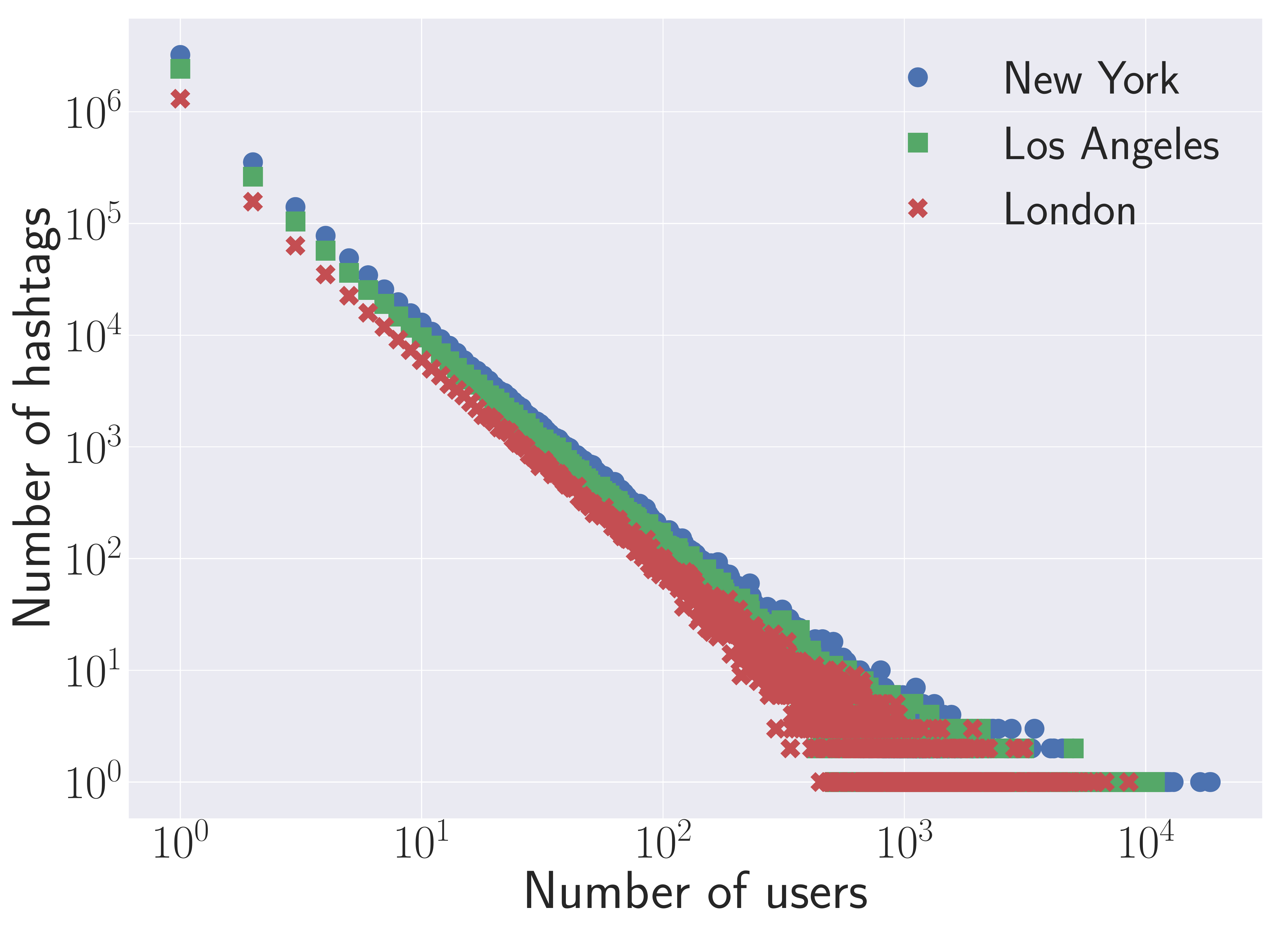}
\caption{}
\label{fig:user_cnt_dist}
\end{subfigure}
\caption{(a) Distribution of the number of times a hashtag is shared in three datasets;
(b) Distribution of the number of users a hashtag is used by in three datasets.
Both x and y axes are in log scale.
}
\end{figure*}

\subsection{Initial Analyses}

\begin{table}[!t]
\centering
\caption{Top 10 hashtags with the highest share times in three datasets.}
\label{table:popular}
\begin{tabular}{ c | c | c}
\toprule
New York & Los Angeles & London\\
\midrule
\#nyc & \#love & \#london \\
\#love & \#losangeles & \#love \\
\#tbt & \#tbt & \#instagood \\
\#nofilter& \#california & \#travel \\
\#brooklyn& \#instagood & \#summer \\
\#latergram& \#nofilter & \#photooftheday \\
\#instagood& \#foodporn & \#food \\
\#art& \#family & \#art \\
\#travel& \#art & \#instadaily\\
\#summer & \#fun & \#architecture \\
\bottomrule
\end{tabular}
\end{table}

\begin{figure*}[!t]
\centering
\begin{subfigure}{\columnwidth}
\centering
\includegraphics[width=0.95\columnwidth]{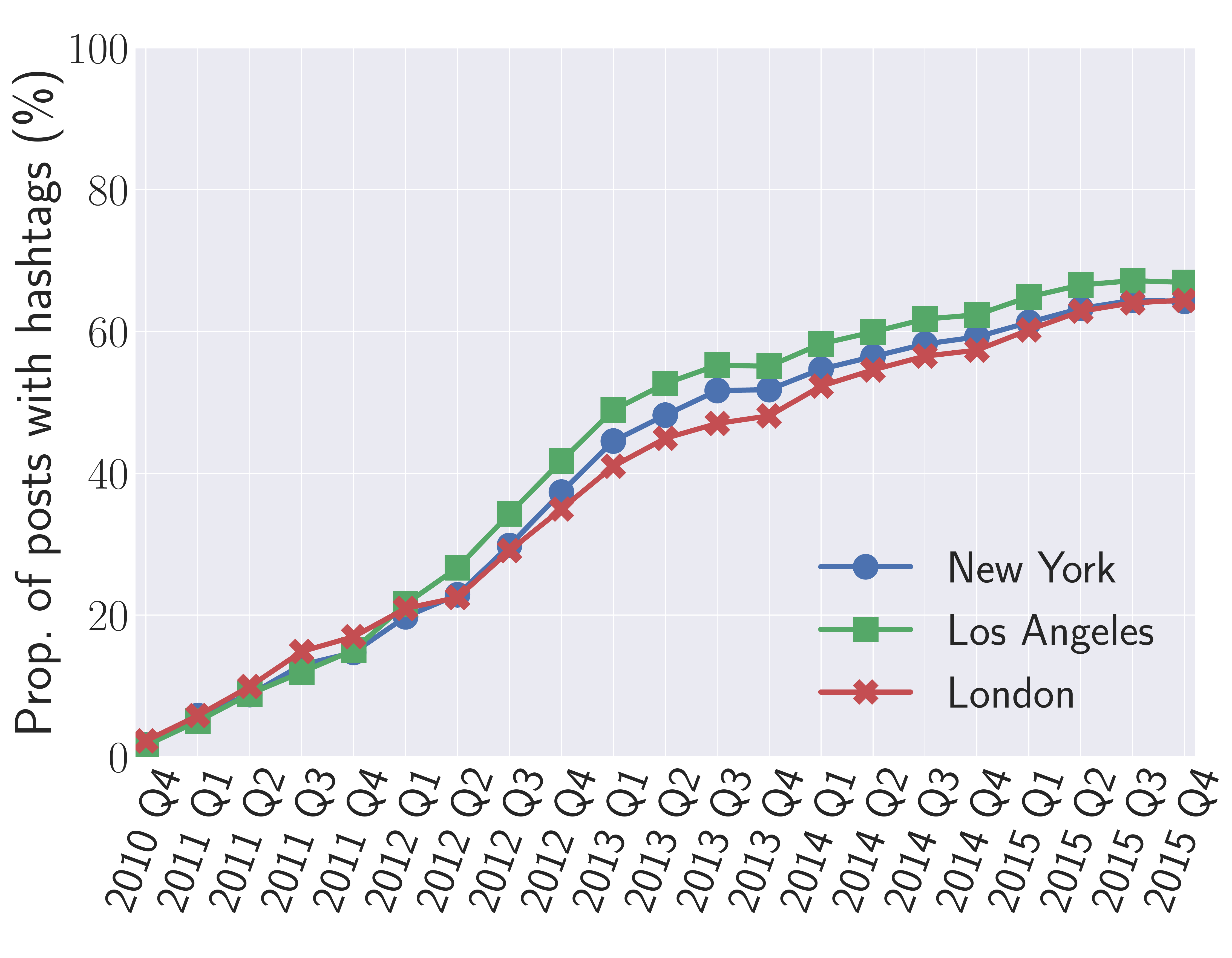}
\caption{}
\label{fig:share_time}
\end{subfigure}
\begin{subfigure}{\columnwidth}
\centering
\includegraphics[width=0.95\columnwidth]{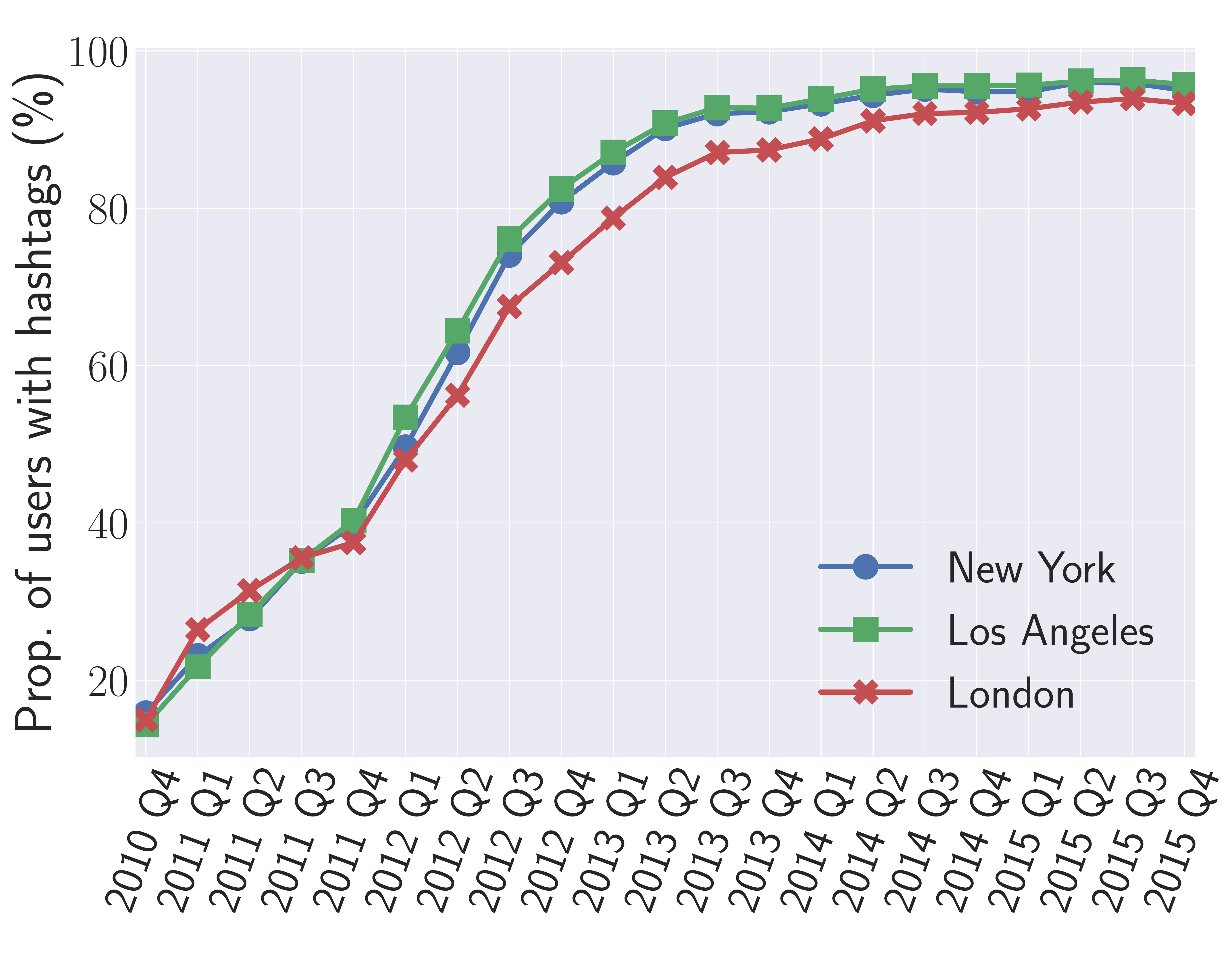}
\caption{}
\label{fig:user_time}
\end{subfigure}
\caption{
(a) Proportion of posts that are associated 
with hashtags from 2010 Q4 to 2015 Q4 in three datasets;
(b) Proportion of users that use hashtags 
from 2010 Q4 to 2015 Q4 in three datasets.
Q1 represents the first quarter of a year.
\label{fig:basic_time}}
\end{figure*} 

Figure~\ref{fig:media_dist} depicts the distribution 
of the number of hashtags in each post.
We observe that more than half of the posts
are associated with at least one hashtag in all the datasets.
Moreover, there are around 30\% posts associated with 1 to 3 hashtags.
The small increase close to 30 is due to the
fact that Instagram imposes an upper bound of 30 hashtags per post.\footnote{\url{https://help.instagram.com/351460621611097}}

Figure~\ref{fig:share_cnt_dist}
plots the distribution of the number of times 
each hashtag is shared, referred to as share times,
while Figure~\ref{fig:user_cnt_dist} plots the distribution
of the number of users each hashtag is used by.
As expected, both distributions follow power law,
i.e., most of the hashtags are shared
only a few times and by a small number of users.
Table~\ref{table:popular}
lists the hashtags with the highest share times.
As our users are sampled by cities,
many of these popular hashtags are related to city names,
such as \#nyc, \#losangeles, and \#london.
However, general popular hashtags are captured as well,
e.g., \#love, \#tbt, and \#instagood.
This indicates that our datasets are suitable for conducting the study.

Figure~\ref{fig:basic_time} depicts hashtags' 
general temporal patterns from 2010 until 2015.
We see that in the fourth quarter of 2010 (2010 Q4) 
when Instagram was launched,
there are less than 2\% of the posts associated with hashtags,
after 5 years (2015 Q4), 
the number becomes almost 70\% (Figure~\ref{fig:share_time}).
Within the same time period, 
the proportion of users that ever use hashtags 
grow from 18\% to 97\%~(Figure~\ref{fig:user_time}).
These results fully demonstrate the popularity of hashtags.

\begin{figure*}
\centering
\begin{subfigure}{\columnwidth}
\centering
\includegraphics[width=0.95\columnwidth]{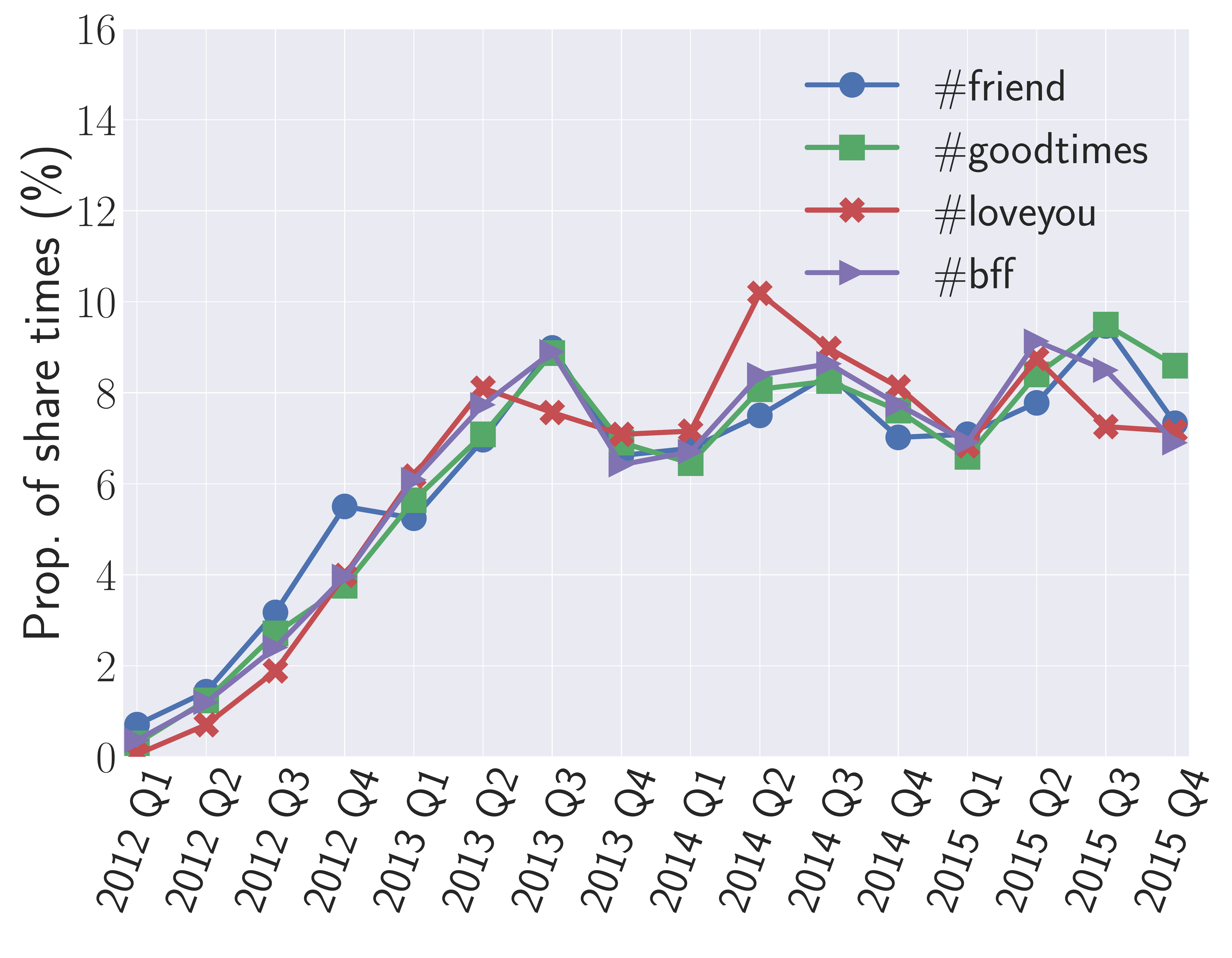}
\caption{Stable}
\label{fig:ny_clu_stable}
\end{subfigure}
\begin{subfigure}{\columnwidth}
\centering
\includegraphics[width=0.95\columnwidth]{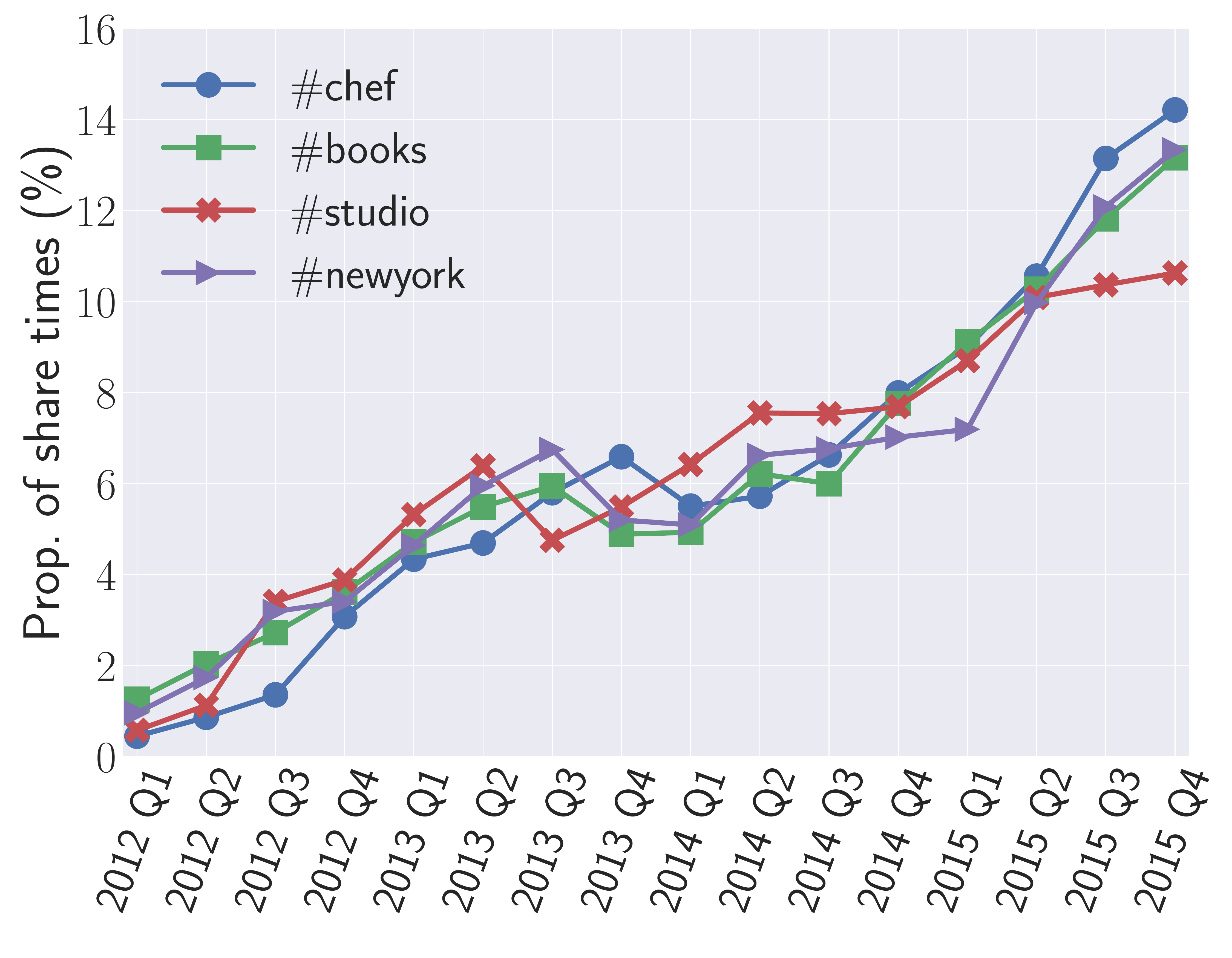}
\caption{Rising}
\label{fig:ny_clu_increase}
\end{subfigure}
\begin{subfigure}{\columnwidth}
\centering
\includegraphics[width=0.95\columnwidth]{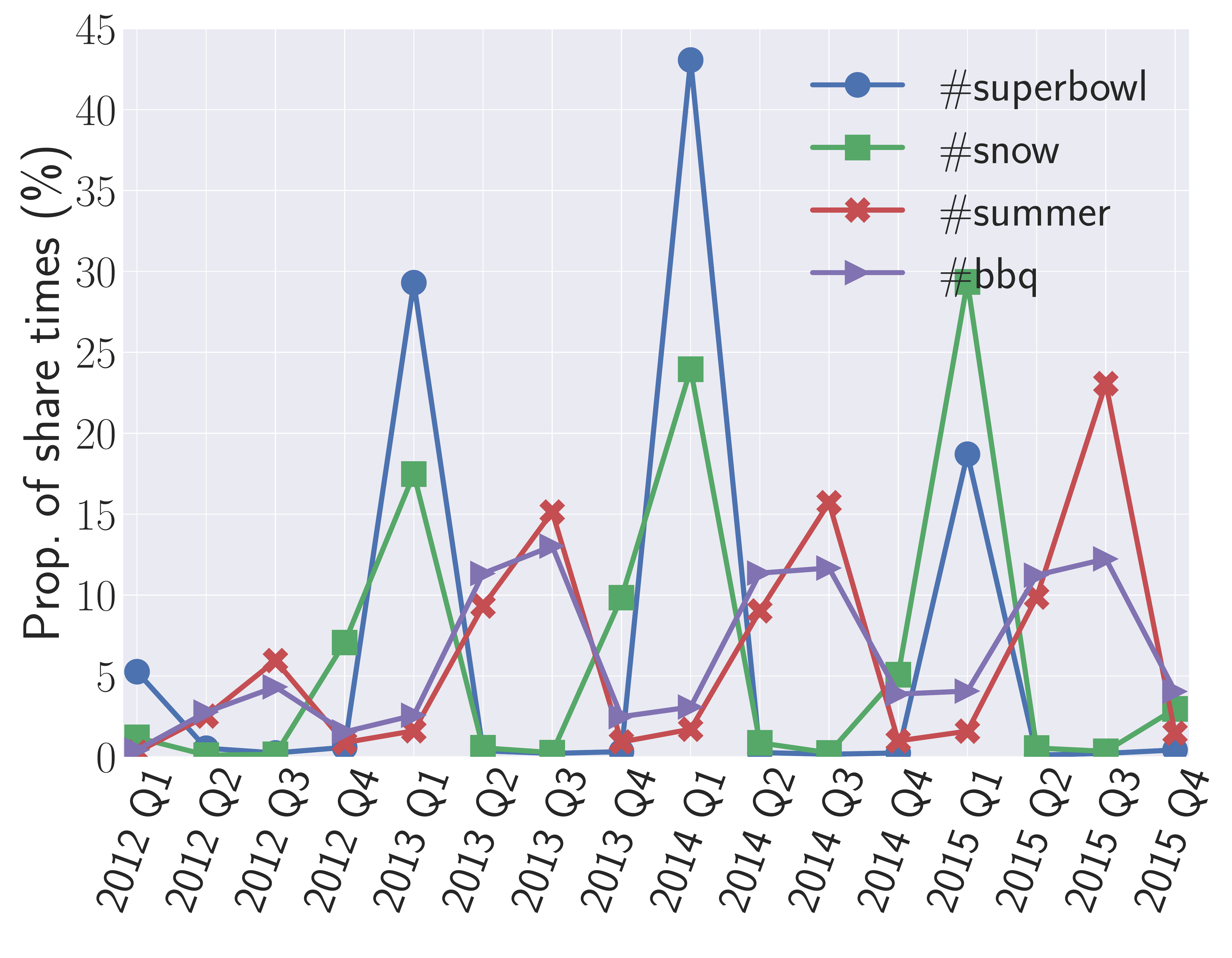}
\caption{Periodic}
\label{fig:ny_clu_periodic}
\end{subfigure}
\begin{subfigure}{\columnwidth}
\centering
\includegraphics[width=0.95\columnwidth]{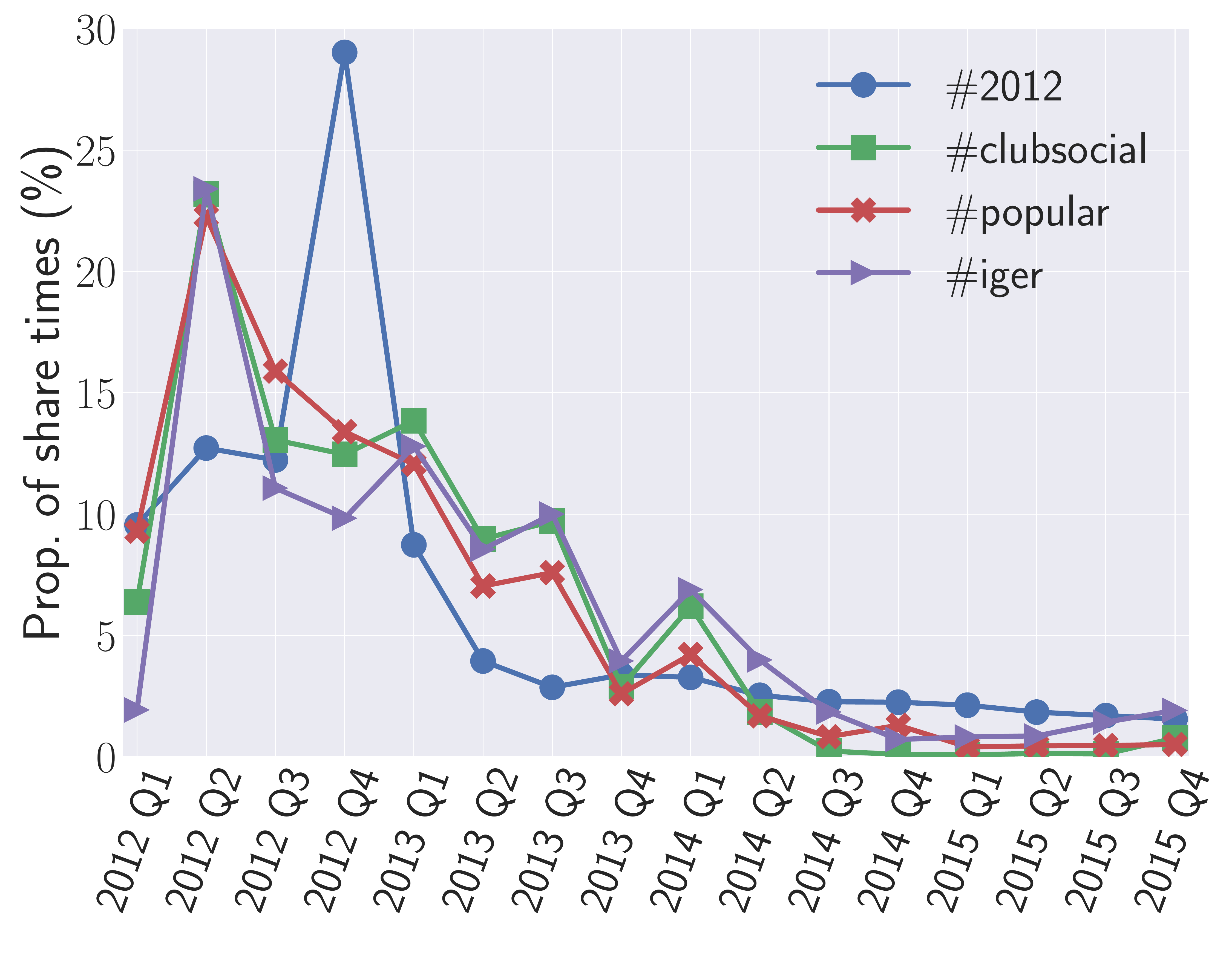}
\caption{Meteor}
\label{fig:ny_clu_decrease}
\end{subfigure}
\caption{Temporal patterns of some example hashtags 
belonging to different temporal clusters
in the New York dataset.}
\label{fig:timeclu}
\end{figure*} 

\section{Hashtags in the Temporal-Spatial Dimension}
\label{sec:spatialtemporal}

This section concentrates on our first research question:
\emph{What are the temporal and spatial patterns of hashtags?}.
We start by investigating the temporal patterns of hashtags, 
then discuss the relation between hashtags and locations.

\subsection{Temporal Patterns}

We hypothesize that different hashtags exhibit different temporal patterns.
Some hashtags should have periodic patterns, such as holiday-related ones,
while others' share times may increase over time.
Besides, there may exhibit other temporal patterns for hashtags.
To perform a quantitative study, we resort to machine learning clustering
to discover different temporal patterns.

We aggregate each hashtag's proportion 
of share times to the granularity of quarters
starting from the first quarter in 2012 
to the last quarter of 2015.\footnote{We neglect the data 
in 2010 Q4 and 2011 due to the small quantity as shown in Figure~\ref{fig:basic_time}).}
This indicates that each hashtag's temporal pattern is organized as a 16-dimension vector (4 years $\times$ 4 quarters).
$k$-means is adopted to perform clustering.
We first try to fit each hashtag's temporal patterns, i.e., the 16-dimension vector,
directly to $k$-means, however, the resulting clusters are not very promising.
This is due to the fact that some hashtags may share similar temporal patterns,
however, their patterns' peak happen at different time, e.g., \#christmas and \#halloween.

Instead, we manually define features over each hashtag's temporal patterns
which are described in Table~\ref{table:temporal}.
To ensure the robustness of our results,
we concentrate on the top 1,000 hashtags in terms of share times~\cite{HLJ16}.
In order to select a suitable number of clusters for $k$-means, i.e., $k$,
we adopt the Silhouette value.
Experimental evaluation suggests 
that $k=4$ leads to the highest Silhouette value (around 0.6)
in all the three datasets.

\begin{table}[!t]
\centering
\caption{Features defined over each hashtag $\hashtag$'s temporal pattern $\temporal_{\hashtag}$.}
\label{table:temporal}
\begin{tabular}{ l }
\toprule
Description\\
\midrule
Standard deviation of $\temporal_{\hashtag}$\\
Largest 3 values in $\temporal_{\hashtag}$\\
Mean of the 3 largest values in $\temporal_{\hashtag}$\\
Standard deviation of the 3 largest values in $\temporal_{\hashtag}$\\
Standard deviation of the 3 largest values' indices in $\temporal_{\hashtag}$\\
Smallest 3 values in $\temporal_{\hashtag}$\\
Mean of the 3 smallest values in $\temporal_{\hashtag}$\\
Standard deviation of the 3 smallest values in $\temporal_{\hashtag}$\\
Standard deviation of the 3 smallest values' indices in $\temporal_{\hashtag}$\\
\bottomrule
\end{tabular}
\end{table}

\begin{figure*}
\centering
\begin{subfigure}{0.69\columnwidth}
\centering
\includegraphics[width=\columnwidth]{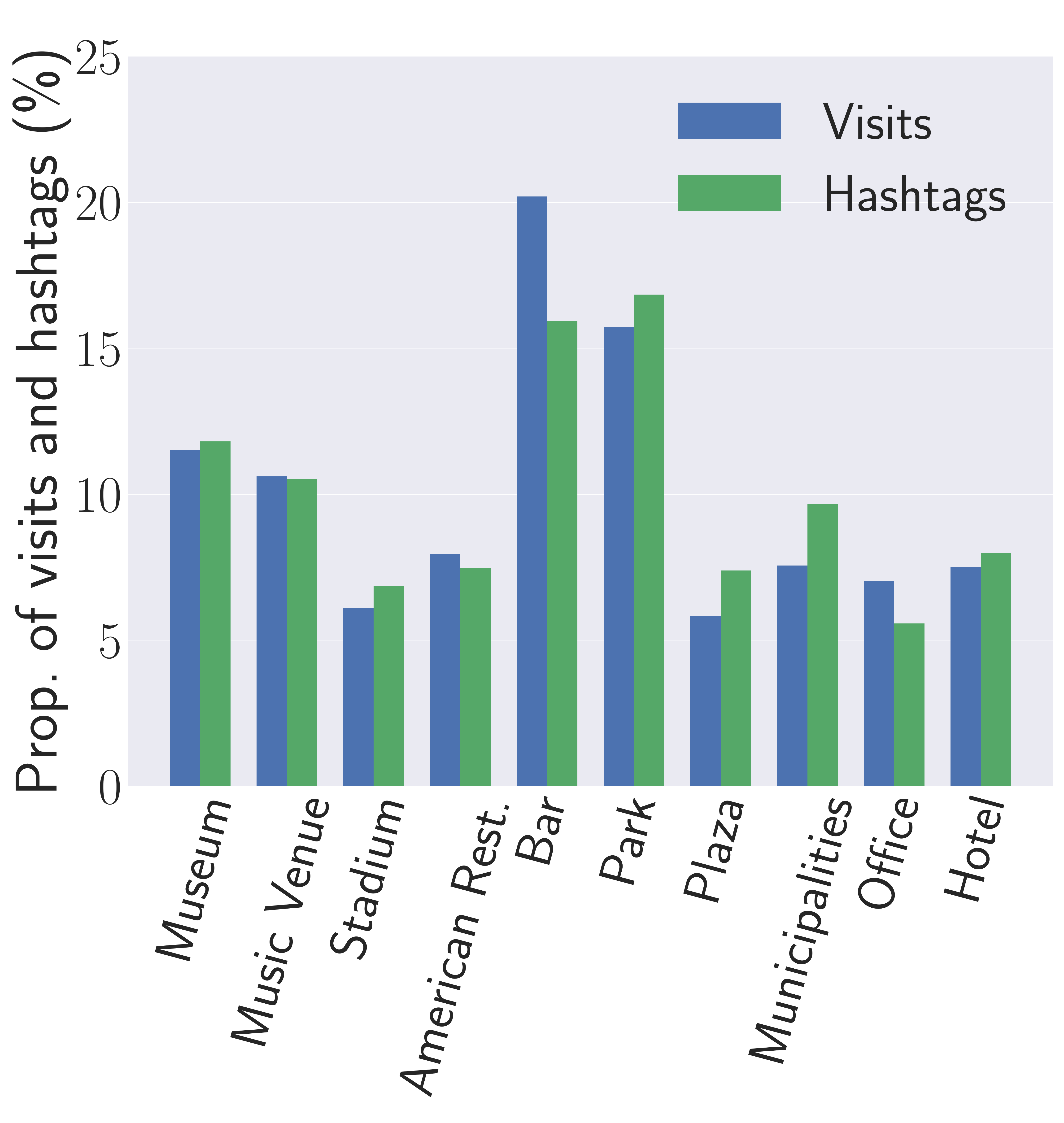}
\caption{New York}
\label{fig:ny_catdist}
\end{subfigure}
\begin{subfigure}{0.69\columnwidth}
\centering
\includegraphics[width=\columnwidth]{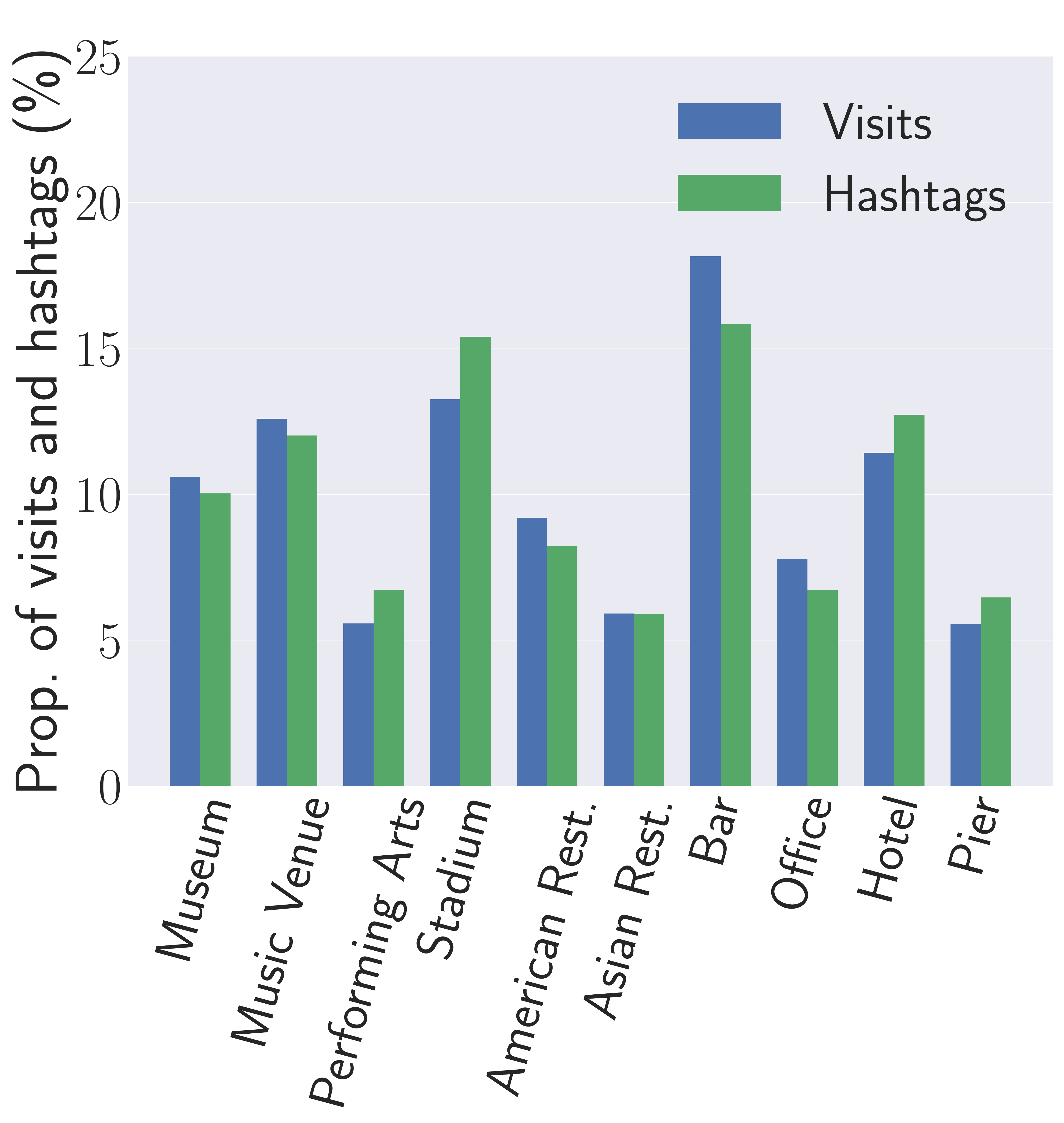}
\caption{Los Angeles}
\label{fig:la_catdist}
\end{subfigure}
\begin{subfigure}{0.69\columnwidth}
\centering
\includegraphics[width=\columnwidth]{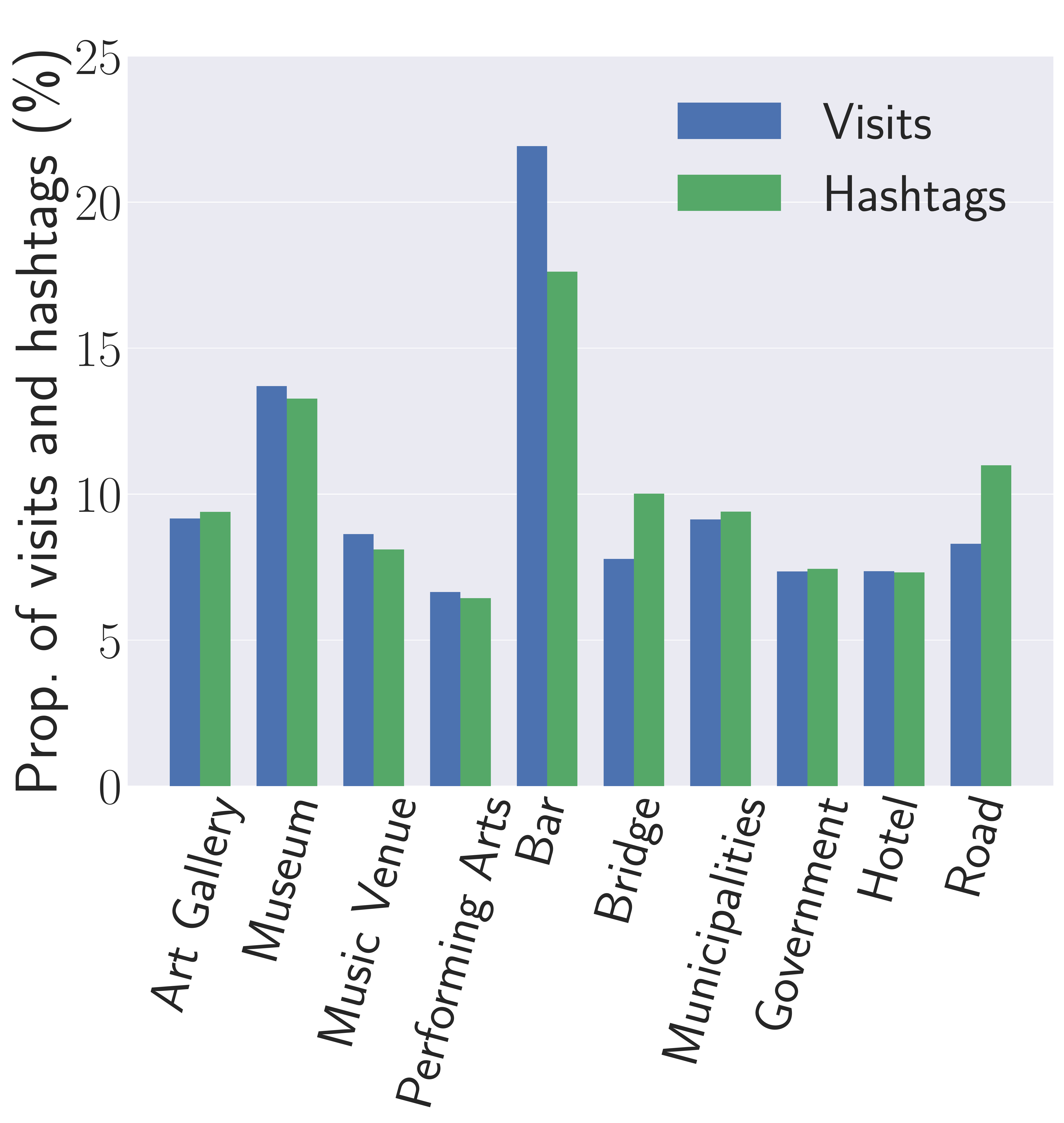}
\caption{London}
\label{fig:london_catdist}
\end{subfigure}
\caption{Proportions of visits and hashtags shared at the most popular location categories 
in three datasets. 
Rest.\ in (a) and (b) means Restaurant.}
\label{fig:catdist}
\end{figure*}

The resulting four clusters are named \emph{Stable}, \emph{Rising},
\emph{Meteor}, and \emph{Periodic}
based on their temporal patterns.
Their detailed descriptions are as follows.
\begin{itemize}
\item Stable: Hashtags in the first cluster 
in the beginning exhibit an increase in their usage.
After a certain time point, their share times become stable.
\item Rising: The share times of hashtags in this cluster increase steadily over time.
\item Meteor: In this cluster, hashtags at some temporal point 
are suddenly shared a large number of times.
\item Periodic: Hashtags in this cluster exhibit periodic popularity.
\end{itemize}
The sizes of the four clusters are not uniform.
The Rising cluster is the largest one containing around 60\% of all hashtags,
followed by the Stable cluster (around 25\%).
Meteor, on the other hand, 
is the smallest cluster with around 7\% of all hashtags.

Figure~\ref{fig:timeclu} depicts the temporal patterns 
of some hashtags in different clusters in the New York dataset.
We can make several interesting observations.
For instance, the Periodic cluster
contains not only hashtags about seasons or holidays, such as \#summer,
but also those about season-related activities, such as \#bbq.
Moreover, hashtags describing (annual) sports events
belong to this cluster as well, e.g., \#superbowl.
As expected, hashtags that are specific to a certain point of time
are categorized in the Meteor cluster,
one example is \#2012.
Interestingly, some hashtags that are specific to Instagram 
also belong to the Meteor cluster.
For instance, \#iger in the second quarter of 2012 gains a large popularity,
while in 2015, no one uses it anymore.

\subsection{Spatial Patterns}

Next, we investigate the spatial patterns of hashtags.
In particular, we are interested in 
at which types of places people are more willing to share hashtags.
To get the type/category of a location in an Instagram post,
we again resort to Foursquare (see Section~\ref{sec:dataset}).
Foursquare organizes all its location categories into 
a two-level tree,\footnote{\url{https://developer.foursquare.com/docs/resources/categories}}
we focus on the fine-grained second-level
which contains more than 300 different location types.

We pick the top 10 location categories 
that users visit the most number of times in each dataset
within the corresponding city,
and calculate the proportion of hashtags shared among these categories.
For comparison,
we further calculate 
the proportion of users' visits to locations under these categories.

Figure~\ref{fig:catdist} plots the results.
First of all, places belonging to the Bar category 
are among the most popular locations for people to visit in all the datasets.
However, the proportions of hashtags shared at bars drop in all the cases,
e.g., the drop in the London dataset is around 5\%.
This means people are less likely to share hashtags at bars.
Another type of locations exhibiting the same result is Office.
On the other hand, people are more willing to share hashtags at outdoor places,
e.g., Park in New York, Pier in Los Angeles, and Road in London.
These results show that the spatial information 
indeed influences users' hashtag-sharing behavior.

\section{Semantic Change}
\label{sec:semantic}

In this section, we address our second research question:
\emph{Do hashtags exhibit semantic displacement?}.
We start by describing how to express each hashtag's semantics,
then focus on semantic displacement.

\subsection{Semantics of Hashtags}

As a new-era language,
hashtags convey interesting meanings.
To study this,
our first step is finding a tool to represent each hashtag's semantics.
Here, we adopt the skip-gram model 
with negative sampling~\cite{MCCD13,MSCCD13}.
Skip-gram, designed following the distributional hypothesis in linguistics,
is essentially a shallow neural network model,
it maps each word into a continuous vector
which preserves the information of 
the word's contexts words in a large corpus.

In our case, we treat each hashtag as a ``word'',
and all hashtags in a post as one ``sentence''.
Then, we perform skip-gram over all the ``sentences'' for each dataset.
We set each hashtag's vector's dimension to 300 
following previous works~\cite{MCCD13,MSCCD13,HLJ16}.

To show the effectiveness of skip-gram 
on capturing hashtags' semantics,
we perform a qualitative study 
to find some hashtags' most semantically similar ones
with respect to the shortest cosine distance between learned vectors.
Table~\ref{table:emb_similar} presents the results 
for \#family, \#sushi, and \#r2d2 in the New York dataset.
As we can see, all the semantically similar hashtags 
found for these hashtags
indeed express quite similar meanings, 
e.g., \#r2d2 is similar to \#c3po, \#droids, and \#bb8.

\begin{table}[!t]
\centering
\caption{10 hashtags that are most semantically similar to \#family, \#sushi, and \#r2d2 
in the New York dataset.}
\label{table:emb_similar}
\begin{tabular}{c | c | c }
\toprule
\#family&\#sushi&\#r2d2\\
\midrule
\#familytime&\#sashimi&\#c3po\\
\#cousins&\#nigiri&\#droids\\
\#aunt&\#sushiporn&\#bb8\\
\#father&\#spicytuna&\#artoo\\
\#mom&\#chirashi&\#astromech\\
\#sisters&\#sushiroll&\#carbonite\\
\#grandparents&\#unagi&\#xwing\\
\#nephews&\#sushirolls&\#obiwankenobi\\
\#familylife&\#shrimptempura&\#lukeskywalker\\
\#siblings&\#californiaroll&\#starwars\\
\bottomrule
\end{tabular}
\end{table}

\subsection{Semantic Displacement}

With the cultural evolution,
some words in natural language exhibit semantic displacement, 
e.g., the word ``gay''.
We are interested in whether semantic displacement can be observed 
on hashtags as well.
Moreover, due to hashtags' inherent dynamic nature,
their semantic displacement should be much faster 
than that of words which normally takes decades,
therefore, our datasets spanning over 5 years 
are sufficient for this study.

To measure semantic displacement of each hashtag,
we first split our datasets by years staring from 2011 to 2015.
Then, we perform skip-gram to map hashtags in each year into a vector.
This means each hashtag has up to 5 vectors.
A \emph{single semantic displacement} of a hashtag
is measured over two consecutive years 
as the cosine distance of the hashtag's two vectors at those years.
The \emph{overall semantic displacement} of a hashtag 
is the mean of all its single semantic displacements.

\begin{table}
\centering
\caption{10 hashtags with the highest overall semantic displacement in three datasets.}
\label{table:top_changed}
\begin{tabular}{c | c | c}
\toprule
New York & Los Angeles & London\\
\midrule
\#ontheroad & \#hashtag & \#iphoneography \\
\#epic & \#epic & \#insta \\
\#ilovemyjob & \#hipstamatic & \#world \\
\#free & \#winning & \#today \\
\#starwars & \#nofilter & \#popular \\
\#winning & \#insta & \#potd \\
\#pictureoftheday & \#ig & \#walking \\
\#nofilter & \#iphone & \#ignation \\
\#iphonography & \#free & \#instamoment \\
\#yolo & \#random & \#studio \\
\bottomrule
\end{tabular}
\end{table}

\begin{table}
\centering
\caption{10 hashtags with the lowest overall semantic displacement in three datasets.}
\label{table:bottom_changed}
\begin{tabular}{c | c | c}
\toprule
New York & Los Angeles & London\\
\midrule
\#nyc & \#sweettooth & \#pet \\
\#beach & \#dessert & \#trees \\
\#sky & \#sweets & \#beach \\
\#sand & \#trees & \#tube \\
\#eastriver & \#palmtrees & \#cat \\
\#leaves & \#soup & \#kitty \\
\#waves & \#ocean & \#skyscraper \\
\#salad & \#sunset & \#thames \\
\#sunset & \#building & \#smoothie \\
\#pasta & \#foodporn & \#glutenfree \\
\bottomrule
\end{tabular}
\end{table}

\begin{figure*}
\centering
\begin{subfigure}{0.69\columnwidth}
\centering
\includegraphics[width=\columnwidth]{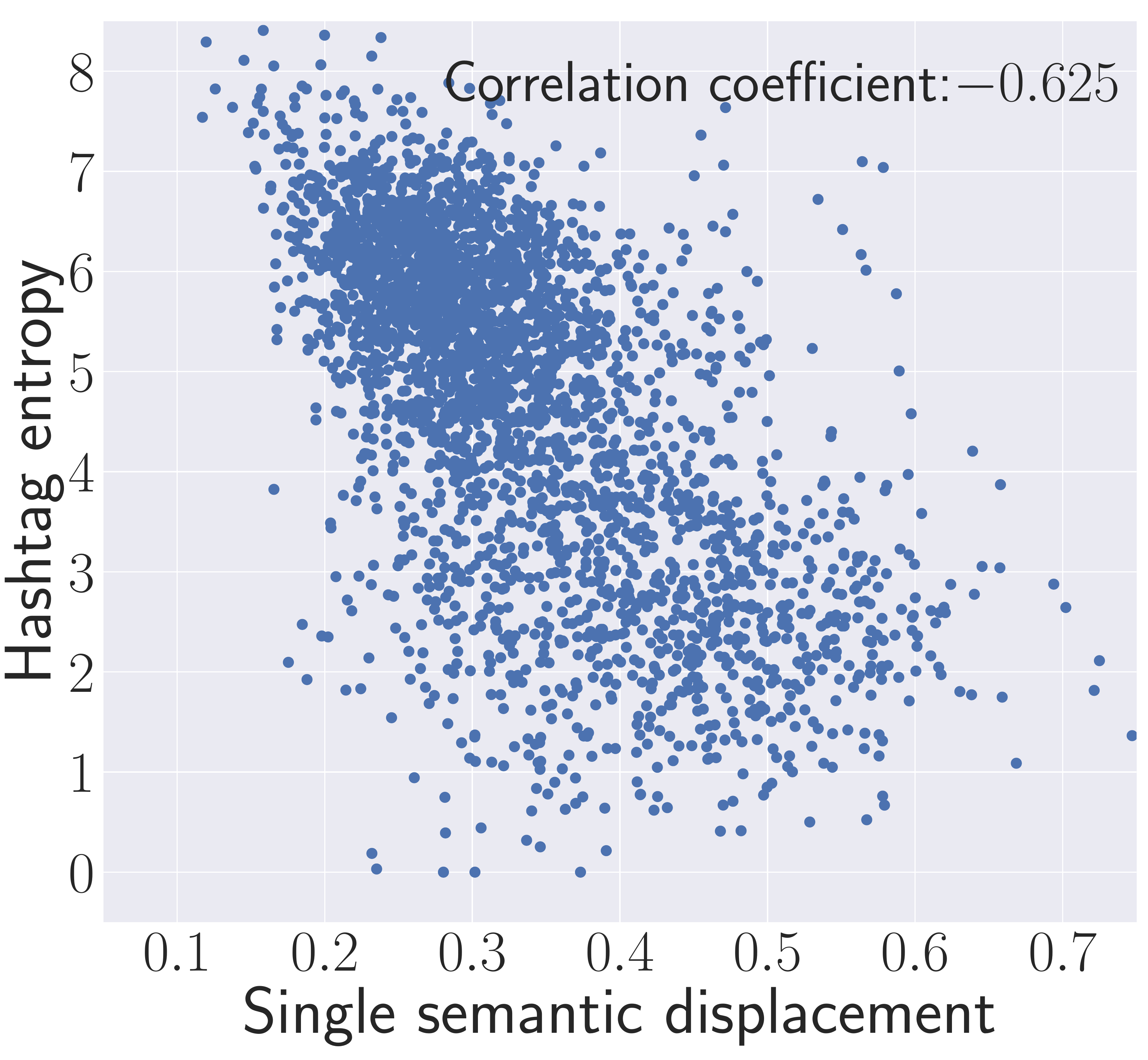}
\caption{New York}
\label{fig:ny_corr}
\end{subfigure}
\begin{subfigure}{0.69\columnwidth}
\centering
\includegraphics[width=\columnwidth]{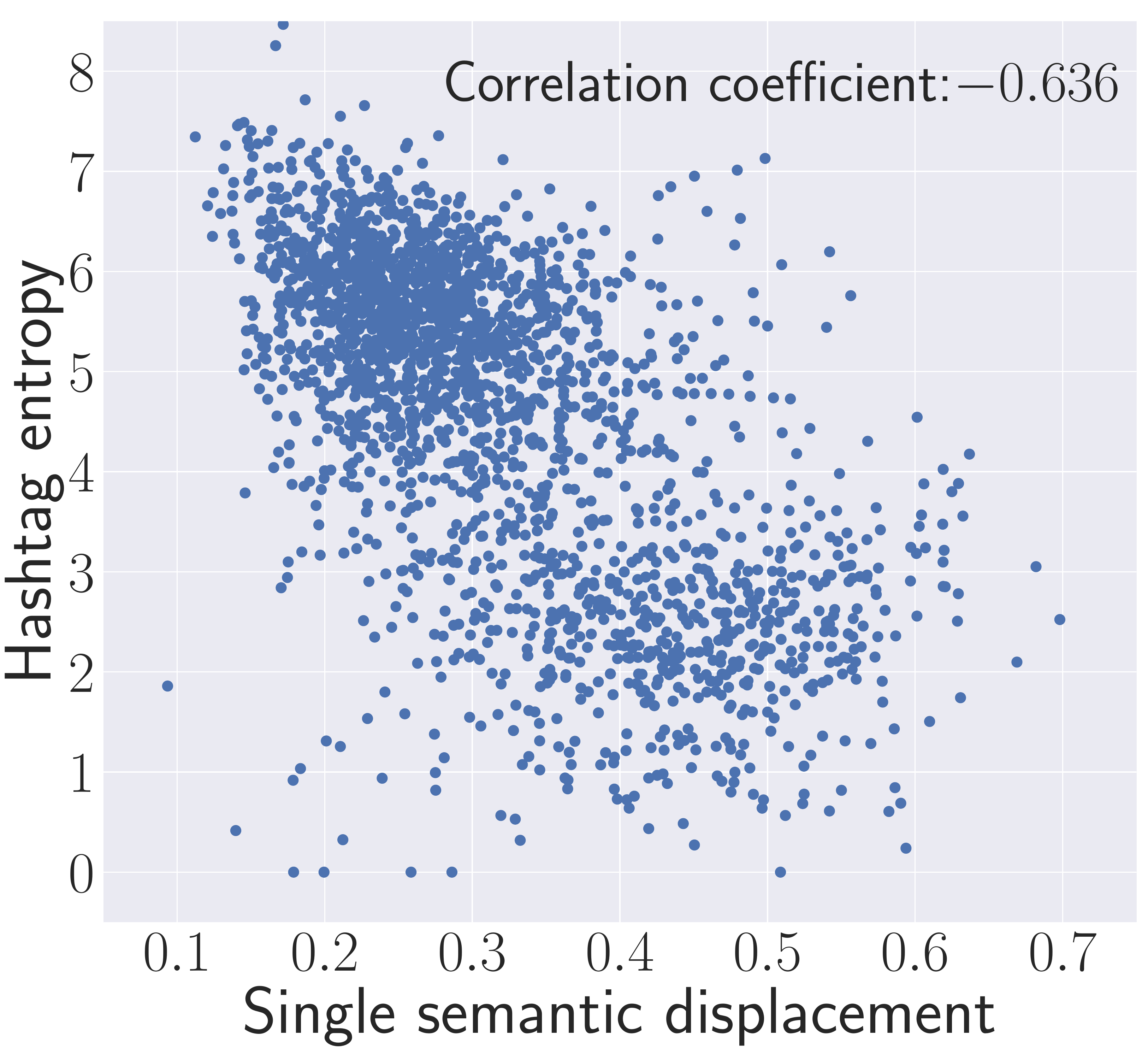}
\caption{Los Angeles}
\label{fig:la_corr}
\end{subfigure}
\begin{subfigure}{0.69\columnwidth}
\centering
\includegraphics[width=\columnwidth]{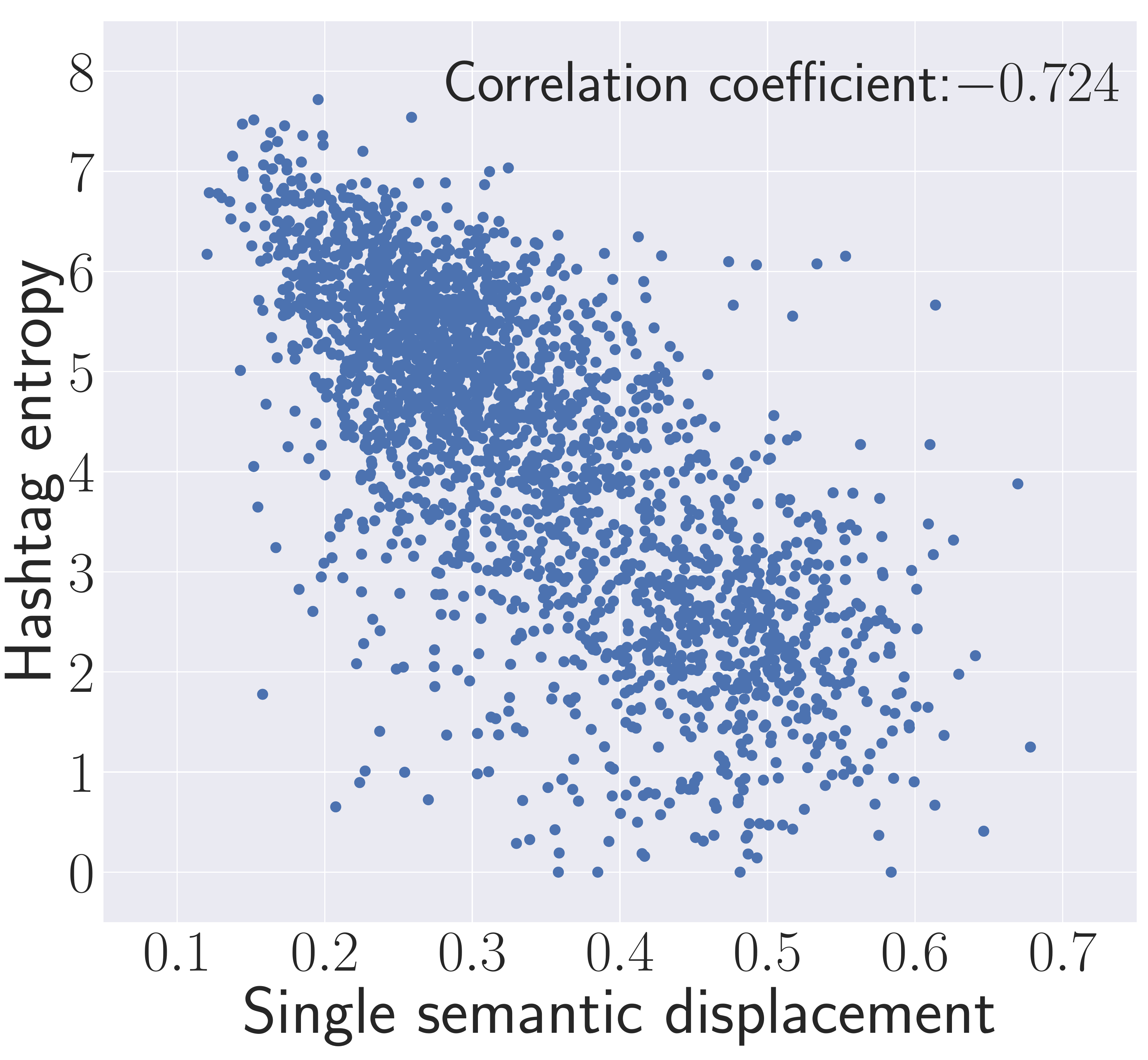}
\caption{London}
\label{fig:london_corr}
\end{subfigure}
\caption{Relation between single semantic displacement 
and hashtag entropy in three datasets.
Hashtag entropy is calculated on a yearly-base.
}
\label{fig:corr}
\end{figure*} 

However, the stochastic nature of the skip-gram model
drives vectors learned at different years to reside at different coordinate axes.
Therefore, 
we need to align each hashtag's vectors at different years together
before calculating their cosine distance~\cite{KAPS15,HLJ16}.
To this end, we apply the orthogonal Procrustes approach 
used by Hamilton et al.~\cite{HLJ16}.
Concretely, we use $\vecmat^{(t)}\in \mathbb{R}^{300\times \vert\Hashtag \vert}$ 
to represent the matrix containing all the hashtags' vectors' at year $t$,
and $\Hashtag$ to represent the set of all hashtags.
Then, we solve the following optimization objective function
\[
\argmin_{\paramat}\vert\vert \paramat\vecmat^{(t)} -\vecmat^{(t+1)}\vert\vert_F \\
\]
\[
\text{subject to } \paramat\paramat^{T} = I
\]
by applying singular value decomposition (SVD).
Here, $\vert\vert \cdot \vert\vert_F$ denotes the Frobenius norm.
After obtaining the result,
we align vectors of a hashtag in two consecutive years together.

Again, to ensure the robustness of our results,
we concentrate on the top 1,000 hashtags that are shared the most number of times.
Our experimental evaluation shows that some hashtags 
indeed exhibit a large semantic displacement.
In our three datasets, there are at least 10\% of the hashtags 
with more than 0.4 overall semantic displacement.
Given the fact that cosine distance 
lies in the range of [0, 2],\footnote{Vectors learned by skip-gram
are in $\mathbb{R}^{300}$, thus, the cosine distance is in [0, 2].}
the semantic shift is pretty large.
Table~\ref{table:top_changed} and Table~\ref{table:bottom_changed}
list the hashtags with the highest and lowest overall semantic displacement.
We see that hashtags that are specific
to Instagram have quite large semantic displacement,
such as \#insta, \#nofilter, and \#ig.
Many of these hashtags are initially created to increase Instagram posts' popularity.
We conjecture that 
when people are more used to these hashtags,
they cannot attract as much attention as before,
thus start to convey different meanings.
On the other hand, we observe from Table~\ref{table:bottom_changed} 
that nature-related hashtags 
are among those whose semantics stay rather stable, e.g., \#sky, \#trees, and \#ocean.
Also, food-related hashtags, such as \#salad, \#soup, and \#smoothie, 
do not change their meanings that much.

Hamilton et al.\ show that high frequency words 
are less likely to change their meanings~\cite{HLJ16}.
However, this is not the case for hashtags.
We perform a correlation analysis between hashtags' frequency, i.e., share times, 
and single semantic displacement, 
and obtain a rather weak correlation (correlation coefficients are around -0.2 in all the datasets).
Instead, we hypothesize that 
hashtags that are more uniformly shared among users 
are less likely to shift their meanings.

We propose \emph{hashtag entropy} to measure 
each hashtag's sharing uniformity among users.
Formally,  a hashtag $\hashtag$, its entropy is defined as:
\[
-\sum_{\user \in \User} p(\user, \hashtag)\log p(\user, \hashtag)
\]
where $\User$ denotes the set containing all the users
and $p(\user, \hashtag)$ represents the proportion of $\hashtag$ shared by user $\user$.
It is worth noting that our hashtag entropy follows the same definition as the Shannon entropy:
Higher entropy implies that the hashtag is more uniformly shared among users.

For each hashtag, we calculate multiple entropy at each year (from 2012 to 2015),
then correlate each entropy with the corresponding single semantic displacement.
Figure~\ref{fig:corr} depicts the results.
In all the three datasets,
we obtain strong negative correlation.
In particular, the correlation coefficient in the London dataset is -0.724.
These results show that 
if a hashtag is shared by many users in a similar frequency, i.e., high entropy,
then it is less prone to semantic displacement.

\section{Hashtags and Friendship}
\label{sec:friends}

In this section, we address our last research question:
\emph{Can hashtags be used to infer social relations?}.
We first describe our methodology on using hashtags for friendship prediction,
then present the experimental evaluation.

\subsection{Methodology}

Social network is the major platform for people to share hashtags,
this naturally leads to the question whether hashtags 
are related to people's social relations.
To answer this question, we perform a friendship prediction task 
solely based on hashtags.

Friendship prediction is normally modeled 
as a binary classification task~\cite{LK07,BL11,SNM11,PSL13,LTZZZZ16,BHPZ17,AB17}.
We can manually define features over two users' hashtags,
and perform prediction either in a supervised manner 
(with the help of machine learning classifier)
or an unsupervised manner.
However, feature engineering is very time-consuming,
and in many cases, the resulting features are not complete.
Also, to define features over two users' hashtags,
a natural approach is concentrating on their common hashtags.
As most pairs of users share no common hashtags,
this approach can be only applied to a small subset of user pairs,
which cannot provide a complete picture
on the relation between hashtags and social networks.

Instead, we choose to learn a hashtag profile, i.e., a feature vector, 
for each user, and compare two users' profiles to perform friendship prediction.
As all users' hashtag profiles (feature vectors) are in the same dimension,
this allows us to predict any pair of users' friendship.

Our hashtag profile learning follows
the recent advancement of graph embedding~\cite{PAS14,TQWZYM15,GL16,BHPZ17,HYL17,
RBL18,MRJTY18,HSLH18,QDMLWT18,QTH18,WZHXGL18,GCHZ18,YYMRHL18}.
In this setting, profiles are automatically learned 
following a general optimization objective function.
Concretely,
we organize users and hashtags into a weighted bipartite graph.
For an edge connecting a user and a hashtag,
its weight equals to the number of times the user shares the hashtag.
We simulate a certain number of random walks
starting from each user, referred to as \emph{walk times}, on the graph.
The transition probability from each node to the next one
follows the corresponding edges' weight.
Each walk has a certain length, specified by the number of steps,
which we refer to as \emph{walk length}.
This leaves us with a set of random walk traces.
Then, we rely on the following optimization objective function 
to learn each user's hashtag profile.
\[
\argmax_\theta\prod_{v\in \User\cup\Hashtag}p(v\vert N(v); \theta)
\]
Here, $N(v)$ denotes the neighborhood of node $v$\footnote{Following 
previous works~\cite{GL16,BHPZ17}, $N(v)$ includes
10 nodes precedent and after $v$ 
in all the random walk traces.}
and $\theta(v)$ is the learned profile of node $v$.
Moreover, $p(v\vert N(v); \theta)$ is modeled with a softmax function.
Different from previous graph embedding approaches
which define their objective functions following skip-gram~\cite{PAS14,GL16},
our objective function is essentially 
the continuous bag-of-words (CBOW) model~\cite{MCCD13,MSCCD13}.
We choose CBOW over skip-gram due to its better performance.
In addition, we apply the negative sampling approach to speed up the learning process.

In the end, for any two users,
we calculate their learned profiles' cosine distance,
and predict them to be friends if the cosine distance 
is below a chosen threshold.\footnote{We also learn 
a set of vectors for all hashtags,
as we concentrate on friendship prediction, 
these hashtag vectors are simply neglected.}
Note that our prediction does not need the knowledge of any existing friendships,
therefore, it is unsupervised.

\subsection{Evaluation}

\begin{figure}
\centering
\includegraphics[width=0.9\columnwidth]{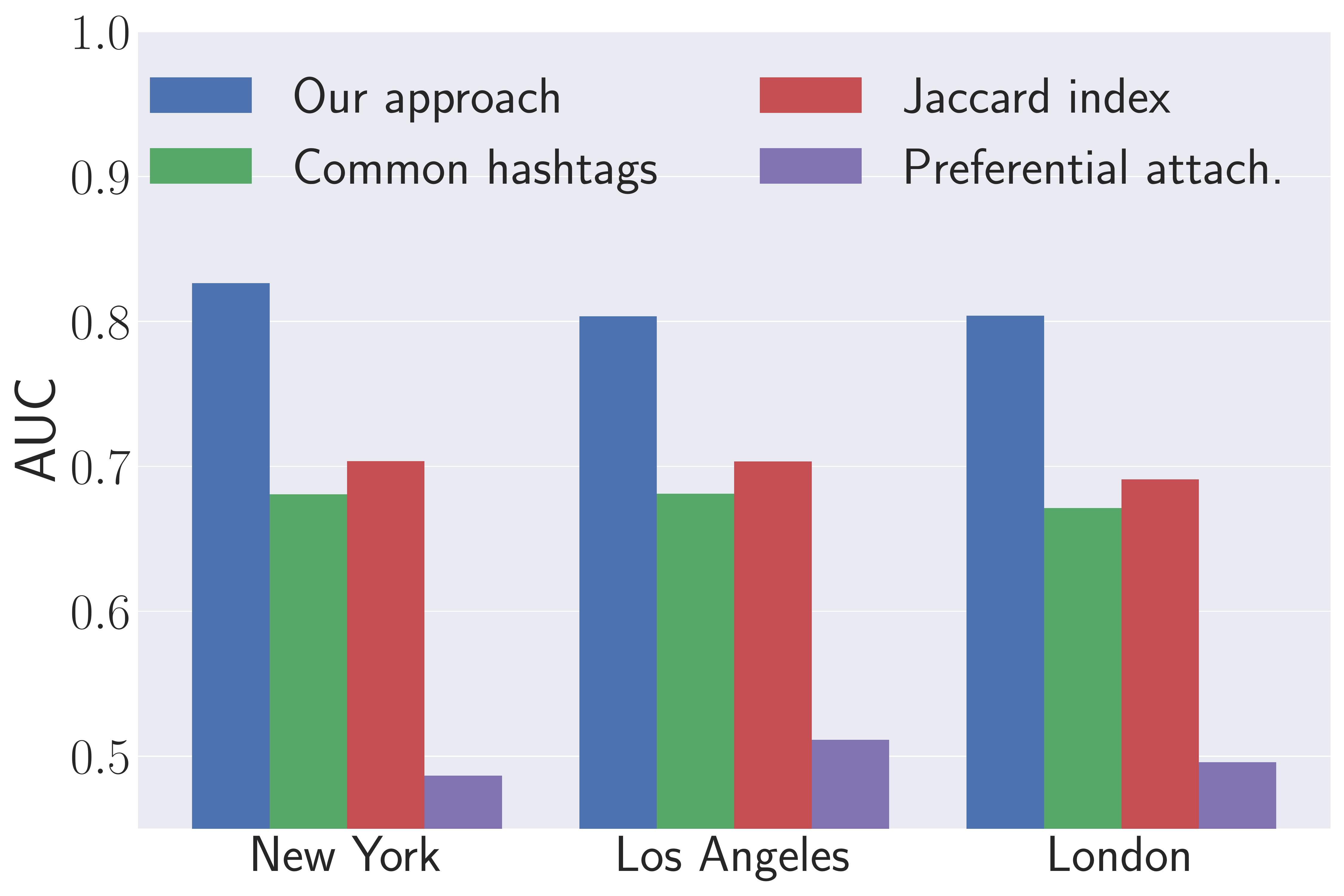}
\caption{AUC for friendship prediction in three datasets. 
Preferential attach. means preferential attachment.}
\label{fig:auc}
\end{figure}

\begin{figure*}
\centering
\begin{subfigure}{0.69\columnwidth}
\centering
\includegraphics[width=\columnwidth]{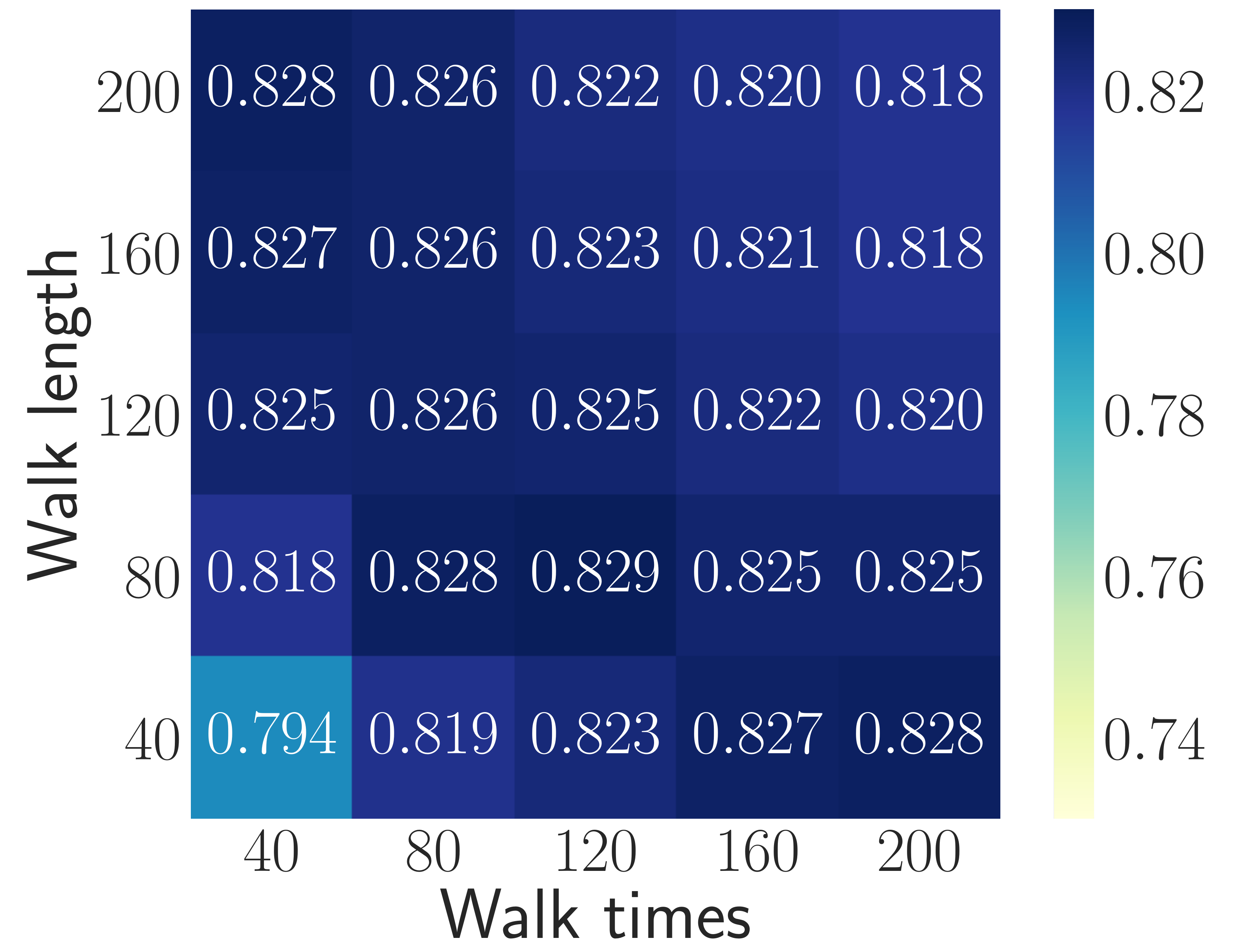}
\caption{New York}
\label{fig:ny_walk_len_walk_times}
\end{subfigure}
\begin{subfigure}{0.69\columnwidth}
\centering
\includegraphics[width=\columnwidth]{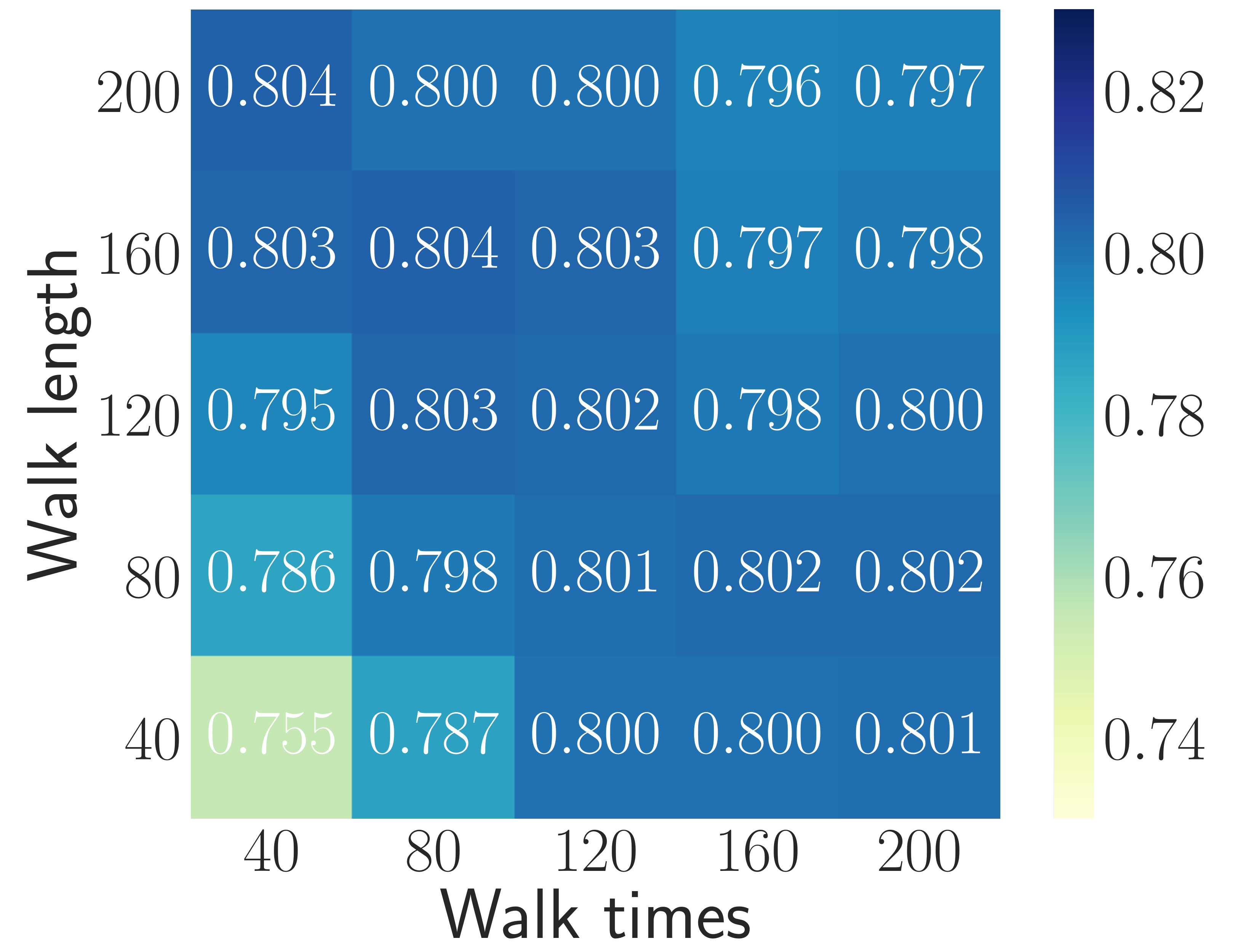}
\caption{Los Angeles}
\label{fig:la_walk_len_walk_times}
\end{subfigure}
\begin{subfigure}{0.69\columnwidth}
\centering
\includegraphics[width=\columnwidth]{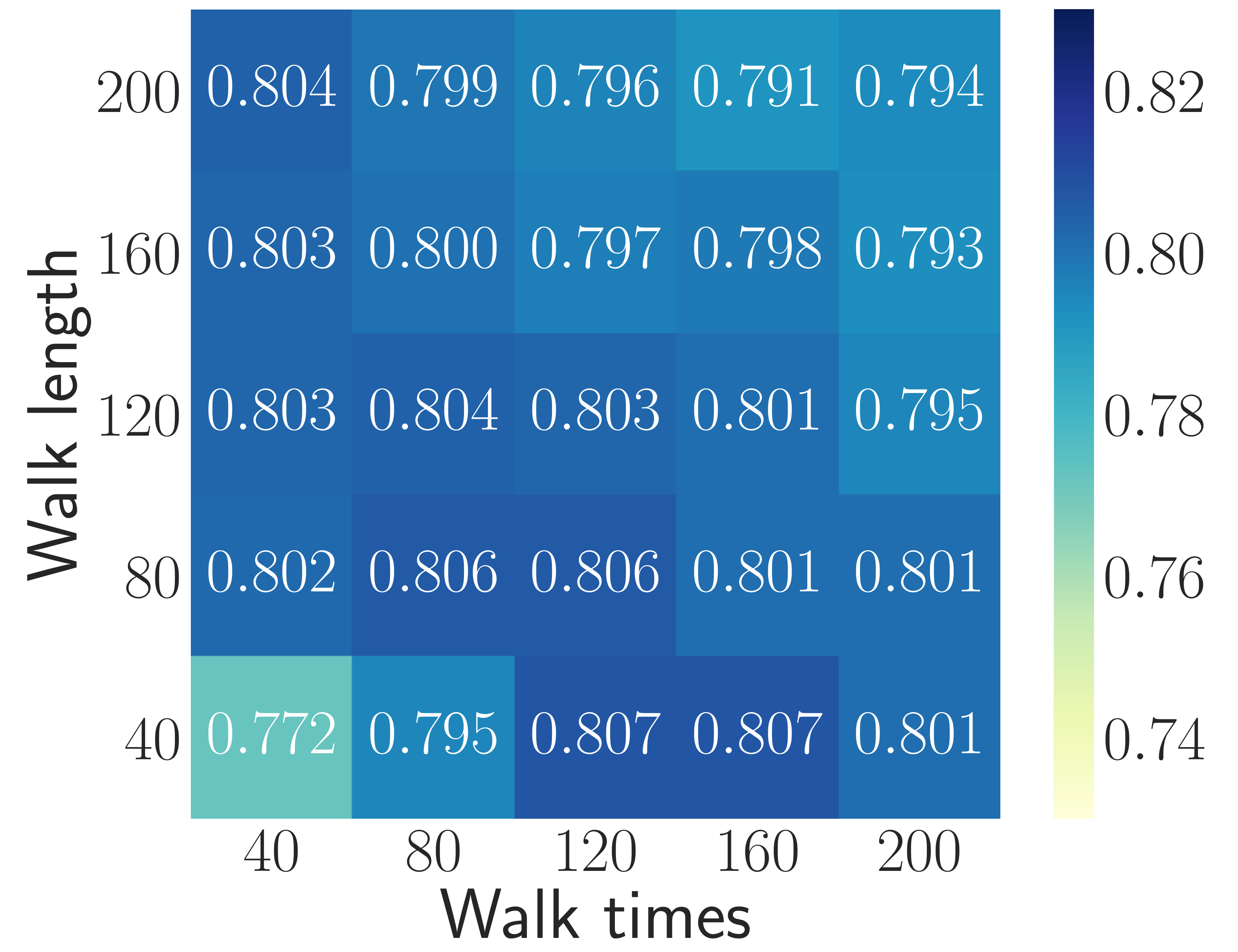}
\caption{London}
\label{fig:london_walk_len_walk_times}
\end{subfigure}
\caption{Influence of walk length and walk times
on the prediction performance in three datasets. 
Each value in the matrix represents the corresponding AUC.}
\label{fig:walk_len_walk_times}
\end{figure*} 

\noindent\textbf{Experimental Setup.}
For each dataset,
we randomly sample the same number of stranger pairs as the number of friend pairs
to construct the negative cases.
Then, we compute cosine distances for all friend and stranger pairs.
AUC (area under the ROC curve)
is adopted as the evaluation metric~\cite{SNM11,BHPZ17,SZHBFB19,HZHBTWB19}.
There are mainly three hyperparameters in our model,
i.e., walk length, walk times, and the dimension of the learned profile.
Based on cross validation, we set them to 120, 80, and 512, respectively.

\begin{table}
\centering
\caption{Formal definition of baseline models for friendship prediction.
$\Hashtag(\user)$ is the set containing all the hashtags $\user$ has shared.}
\label{table:baseline}
\begin{tabular}{l | c}
\toprule
Baseline & Definition\\
\midrule
Common hashtags & $\vert\Hashtag(\user)\cap\Hashtag(\user')\vert$\\
Jaccard index & $\frac{\vert\Hashtag(\user)\cap\Hashtag(\user')\vert}{\vert\Hashtag(\user)\cup\Hashtag(\user')\vert}$\\
Preferential attachment & $\vert\Hashtag(\user) \vert\cdot \vert\Hashtag(\user') \vert$\\
\bottomrule
\end{tabular}
\end{table}

\medskip
\noindent\textbf{Baseline Models.}
We further establish three baseline models following 
the traditional link prediction approach~\cite{LK07},
namely 
Jaccard index,
common hashtags,
and preferential attachment.
Their formal definition is presented in Table~\ref{table:baseline}.

\medskip
\noindent\textbf{Results.}
Figure~\ref{fig:auc} depicts the AUCs for friendship prediction.
In all the three datasets, we obtain more than 0.8 AUC
which shows that hashtags can provide strong signals on inferring social relations.
In particular, the prediction in the New York dataset achieves the best performance.

Our approach also outperforms all the three baseline models significantly.
For the best baseline model, i.e., Jaccard index, 
we achieve around 20\% performance gain in all the datasets.
One of the major advantages of our approach
is its ability to predict two users' friendship 
regardless of whether they share common hashtags.
We observe that even for pairs of users sharing no common hashtags, 
our prediction still achieves a decent performance in all the datasets:
0.759 AUC in New York, 0.733 AUC in Los Angeles, and 0.728 AUC in London.
This further demonstrates the effectiveness of our approach.

We also study the influence of hyperparameters
on the prediction performance.
Figure~\ref{fig:walk_len_walk_times}
depicts the heatmap when jointly tuning walk times and walk length.
We see that when these two values are small, i.e., 40,
the prediction results are relatively weak in all the three datasets.
When increasing both, the performance gets better.
However, the relation between the prediction result 
and the magnitude of the hyperparameters is not monotonic:
When we set both walk times and walk length to 200,
the prediction results drop in all cases.
Note that we also perform the same study 
on the dimension of each learned profile,
and observe that 512 leads to the best prediction.

\section{Related Work}
\label{sec:related}

Hashtags provide us with an unprecedented chance 
to understand the modern society.
Researchers have studied hashtags 
from many perspectives~\cite{DTR10,RMK11,RCMGFM11,LWTL11,CRFGFM11,
CZLMZ12,YSZM12,MWR12,LGRC12,TR12,MSC12,KHLZ12,GLQ13,GOAJ13,PWSS13,LMKBL13,
OCPJWMA14,QSA14,KMFYZ14,SCFYCQA15,JHSL15,FPS15,AW16,GMGM16,
MAH16,OWG16,CTYYL16,OAWHT17,BFL18,ZHRLPB18}.

Souza et al.\ use \#selfie
to study the phenomenal self-portrait behavior in OSNs~\cite{SCFYCQA15}.
They collect a large dataset from Instagram,
and show that the amount of posts associated with \#selfie
increases 900 times from 2012 to 2014.
Moreover, their results suggest that posts with \#selfie
attract more likes and comments than others.
The authors further show that there exist cultural variations
of selfie behavior across countries.

Mejova et al.\ study food-related hashtags, such as \#foodporn, 
on Instagram
to understand people's dining behavior on a global scale~\cite{MAH16}.
They first show that desserts in social media are dominating over local cuisines.
Then, they discover through hashtags
that food can motivate people to engage in a healthier life style.
Similar to \#selfie,
Mejova et al.\ show that posts associated with healthy-related hashtags 
attract more likes than others.

David et al.\ use hashtags together with the smiley face emoji
to perform sentimental analysis on Twitter~\cite{DTR10}.
In this work, hashtags are used as the sentimental labels,
and the authors summarize features over tweets.
Evaluation results show that
their classification achieves very effective performance.

Multiple works also use hashtags to study various political movements.
Olteanu et al.\ analyze the demographics behind \#blacklivesmatter,
and conclude that African-Americans and young females 
engage more with the hashtag than others~\cite{OWG16}.
Manikonda et al.\ perform a comparative analysis of \#metoo
shared on Twitter and Reddit~\cite{MBLK18}.
They observe that posts on Reddit concentrate on sharing personal stories 
while tweets express public support for the \#metoo movement.

Zhang et al.\ study the privacy implications of sharing hashtags~\cite{ZHRLPB18}.
They first utilize a simple bag-of-words model and a random forest classifier 
to perform location prediction based on hashtags.
Evaluation shows that their approach
achieves more than 70\% accuracy over fine-grained locations.
Then, the authors propose a privacy-preserving system, namely Tagvisor,
to mitigate the location privacy risks.
Tagvisor implements three different obfuscation mechanisms including hiding, replacement, and (location category) generalization.
Extensive experiments suggest that by obfuscating two hashtags,
Tagvisor can successfully mislead the location prediction model,
while maintaining a high-level utility with respect to hashtag semantics.

While the above works concentrate on either a certain type of hashtags or a certain property of hashtags,
Ferragina et al.\ perform a general analysis of hashtags on Twitter~\cite{FPS15}.
In particular, they concentrate on the semantics of Twitter's hashtags.
They first build a hashtag-entity graph over a large number of tweets.
Then, the authors perform two natural language processing tasks, 
namely hashtag relatedness and hashtag classification,
based on the features extracted from the constructed hashtag-entity graph.
Experimental results show that this approach outperforms state-of-the-art solutions by a large extent.
Besides targeting a different OSN than Ferragina et al.~\cite{FPS15}, 
we perform a much broader analysis on hashtags with several new angles,
such as spatial pattern, semantic displacement, and social signal.

\section{Conclusion}
\label{sec:conclu}

In this paper,
we perform the first large-scale analysis 
on understanding hashtags shared on Instagram.
Our study is centered around three research questions
which aim at understanding hashtags from three different dimensions,
i.e., the temporal-spatial dimension, the semantic dimension,
and the social dimension.
We collect three large datasets from Instagram 
containing more than 7 million hashtags shared over 5 years
to perform our analyses.

We first show that hashtags can be categorized 
into four different clusters according to their temporal patterns,
and people are more willing to share hashtags at certain places, such as parks.
We then discover that some hashtags indeed exhibit large semantic displacement.
Moreover, we propose a notion namely hashtag entropy
and show the strong negative correlation 
between hashtag entropy and semantic displacement.
In the end, we propose a bipartite graph embedding model
to summarize users' hashtag profiles
and rely on these profiles to perform friendship prediction.
The effective prediction performance suggests
that there is a strong connection between hashtags and social relations.

\section{Acknowledgments}

This work was partially supported by the German Federal Ministry of Education and Research (BMBF) through funding for the Center for IT-Security, Privacy and Accountability (CISPA) (FKZ: 16KIS0656). The author would like to thank Dr. Sandra Strohbach and Dr. Mathias Humbert for their constructive feedback on the manuscript.

\bibliographystyle{ACM-Reference-Format}
\bibliography{normal_generated}

\end{document}